\documentclass[twocolumn,superscriptaddress,amsmath,amssymb,aps,prb,longbibliography]{revtex4-1}

\usepackage{graphicx}
\usepackage{dcolumn}
\usepackage{bm}
\usepackage[colorlinks=true,linkcolor=red]{hyperref}
\usepackage{xfrac}
\usepackage{physics}
\usepackage[normalem]{ulem}
\usepackage{xcolor}

\newcommand{\pcsadd}{Center for Theoretical Physics of Complex Systems, Institute for Basic Science(IBS), Daejeon 34126, Korea}
\newcommand{\ustadd}{Basic Science Program(IBS School), Korea University of Science and Technology(UST), Daejeon 34113, Korea}

\newcommand{\mh}{\mathcal{H}}
\newcommand{\efb}{\ensuremath{E_\text{FB}}}
\newcommand{\pt}{\ensuremath{\mathcal{PT}}}
\newcommand{\pts}{{\pt}-symmetry}
\newcommand{\vpsi}{\vec{\psi}}
\newcommand{\bpsi}[1]{\bra{\psi_{#1}}}
\newcommand{\kpsi}[1]{\ket{\psi_{#1}}}

\newcommand{\PCLS}{\Psi_\text{CLS}}

\begin{document}

\title{Non-Hermitian flatband generator in one dimension}

\author{Wulayimu Maimaiti}
\affiliation{\pcsadd}
\affiliation{\ustadd}

\author{Alexei Andreanov}
\affiliation{\pcsadd}
\affiliation{\ustadd}

\date{\today}

\begin{abstract}
    Dispersionless bands -- flatbands -- have been actively studied thanks to their interesting properties and sensitivity to perturbations, which makes them natural candidates for exotic states. In parallel non-Hermitian systems have attracted much attention in the recent years as a simplified description of open system with gain or loss motivated by potential applications. In particular, non-Hermitian system with dispersionless energy bands in their spectrum have been studied theoretically and experimentally. Flatbands require in general fine-tuning of Hamiltonian or protection by a symmetry. A number of methods was introduced to construct non-Hermitian flatbands relying either on a presence of a symmetry, or specific frustrated geometries, often inspired by Hermitian models. We discuss a systematic method of construction of non-Hermitian flatbands using 1D two band tight-binding networks as an example, extending the methods used to construct systematically Hermitian flatbands. We show that the non-Hermitian case admits fine-tuned, non-symmetry protected flatbands and provides more types of flatbands than the Hermitian case.
\end{abstract}

\maketitle

\section{Introduction}
\label{sec:introduction}

Macroscopically degenerate flatbands (FB) -- dispersionless energy bands in translational invariant tight-binding networks -- have been an active area of research due to their sensitivity to perturbations and interactions, that results in quantum phases with interesting properties.~\cite{maksymenko2012flatband,leykam2018artificial,leykam2018perspective} Since in general FB require fine-tuning of the Hamiltonian parameters many works have explored the construction of flatband Hamiltonians.~\cite{maimaiti2017compact,toikka2018necessary,maimaiti2019universal,maimaiti2020thesis,
flach2014detangling,maimaiti2020flatband,ramachandran2017chiral,mielke1991ferromagnetism,mielke1991ferromagnetic,
mielke1993ferromagnetism,tasaki1992ferromagnetism,tasaki2008hubbard,tasaki1998from} Flatbands have been experimentally realized in various Hermitian settings: photonic systems,~\cite{guzman2014experimental,vicencio2015observation,mukherjee2015observation} ultra-cold atoms,~\cite{jo2012ultracold} exciton-polariton condensates.~\cite{masumoto2012exciton} The effects of various perturbations on FBs were studied in many contexts: itinerant ferromagnetism,~\cite{mielke1991ferromagnetism,mielke1992exact,mielke1993ferromagnetism,tasaki1992ferromagnetism,tasaki2008hubbard}  disorder,~\cite{flach2014detangling,leykam2017localization,bodyfelt2014flatbands,danieli2015flatband,chalker2010anderson,goda2006inverse,nishino2007flat,chen2017disorder}  interactions,~\cite{tovmasyan2018preformed,peotta2015superfluidity,julku2016geometric,tovmasyan2016effective} external fields,~\cite{khomeriki2016landau,kolovsky2018topological} nonlinearity.~\cite{danieli2018compact,johansson2015compactification,real2018controlled,belicev2017localized,perchikov2017flat}

Recently non-Hermitian systems~\cite{moiseyev2011nonhermitian,bagarello2015nonselfadjoint} have attracted attention in different areas of physics due to their exciting properties, such as complex spectrum, non-orthogonal eigenstates,~\cite{curtright2007biorghogonal,brody2013biorthogonal} exceptional points.~\cite{heiss2004exceptional,heiss2001chirality,berry2004nonhermitian,hern2006nonhermitian,gao2015observation,jinhui2014nhdegeneracy}  Moreover non-Hermitian Hamiltonians can account for coupling with environment in open quantum systems and simplify the analysis reducing the large number of degrees of freedom associated with the environment.~\cite{eleuch2018lossgain,rotter2009nonhermitian,rotter2018equilibrium} After the discovery that parity-time ($\pt$)-symmetric non-Hermitian Hamiltonians have real eigenvalues,~\cite{bender1998realspectra,bender2007makingsense,znojil1999nhharmonic} $\pt$ symmetry has been realized in optical systems with gain/loss,~\cite{makris2008beam} leading to an explosion of studies in $\pt$-symmetric non-Hermitian photonics.~\cite{ganainy2007theory,musslimani2008optical,klaiman2008visualization,guo2009observation,ruter2010observation,kottos2010broken,wimmer2015opticalsoliton,feng2017nhphotonics,teimourpour2017rubustness,zhang2018nhoptics,ramy2019dawn} Recently non-Hermitian systems with nontrivial topological properties started to attract attention.~\cite{leykam2017edgemodes,gong2018topological,yuce2015topological,zeuner2015topological}

As the interest to non-Hermitian systems is increasing, flatbands have been predicted in various non-Hermitian systems~\cite{chern2015pt,leykam2017flat,qi2018defect,zyuzin2018flat,ge2018non,zhang2020nonhermitian}, including optical lattices using destructive interference~\cite{zhang2020flatband}, where compact localized states are present. Non-hermitian flatbands have also been realized experimentally.~\cite{tobias2019experimental} Non-Hermitian flatbands (NHFB) exhibit many interesting phenomena such as polynomial power increase of flatband eigenstates~\cite{ge2018non}, manipulation of light localization.~\cite{ramezani2017non} In all of these studies the existence of non-Hermitian flatbands was a consequence of either an existing Hermitian flatband,  a specific frustrated geometry, or a presence of a specific symmetry in the model, such as \pts,~\cite{ramezani2017non,leykam2017flat} or non-Hermitian particle-hole symmetry.~\cite{ge2018non,qi2018defect} An important question is whether it is possible to construct NHFBs from fine-tuning systematically, similarly to the Hermitian case.~\cite{maimaiti2017compact,maimaiti2019universal} The examples of Hermitian flatbands strongly suggest that the NHFB should exist in generic non-Hermitian system without protection by any specific symmetry, merely due to fine-tuning. It is also natural to ask whether such generic, non-symmetry protected NHFB have similar properties to their Hermitian counterparts, like existence of compact localized states (CLS), etc. It is therefore important to devise a systematic construction and classification of non-Hermitian flatbands, by analogy with  the Hermitian case.~\cite{maimaiti2017compact,maimaiti2019universal,maimaiti2020flatband}

In this work we report a systematic definition and construction of non-Hermitian flatband Hamiltonians in 1D. We focus on the case of networks with only two bands by analogy with our previous work on systematic classification of Hermitian flatbands.~\cite{maimaiti2017compact} We introduce in Sections~\ref{sec:def}-\ref{sec:gen}, the classification of NHFB: perfectly flat, partially flat - either real or imaginary part of the band is flat, or the modulus of the eigenenergy is flat. Based on this classification we explore the various NHFB models and discuss the extension of the generic approach for Hermitian system from Refs.~\onlinecite{maimaiti2017compact,maimaiti2019universal} in Sec.~\ref{sec:res} followed by conclusions.

\section{Setting the stage}
\label{sec:def}

We follow the notation of our previous works, Ref.~\onlinecite{maimaiti2017compact,maimaiti2019universal}, which we adapt to the non-Hermitian case. We consider a one-dimensional ($d=1$) translational invariant non-Hermitian lattice with $\nu=2$ sites per unit cell. We group the amplitudes of a wavefunction $\Psi$ by unit cells: $\Psi=(\vpsi_1,\cdots,\vpsi_i,\cdots)$ where $\vpsi_i$ is a two component vector. We use the two notations - $\vpsi_i$ and $\ket{\psi_i}$ - interchangeably throughout the text. For convenience we only consider nearest neighbor hopping between the unit cells. In the Hermitian case the Hamiltonian is described by the hopping matrices between unit cells: $H_0, H_1$. The intracell hopping matrix $H_0$ is Hermitian and the intercell hopping matrix $H_1$ describes hopping to the left unit cell, while $H_1^\dagger$ describes hopping to the right unit cell. However, in the non-Hermitian case $H_0$ is no longer Hermitian and hoppings to the right and to the left neighbor unit cells need not be Hermitian conjugates of one another. Therefore we replace $H_1\to H_l$ and $H_1^\dagger\to H_r$. Then the eigenvalue equation giving the spectrum of the chain reads
\begin{gather}
    \label{eq:eig-nh}
    H_0 \vpsi_n + H_r \vpsi_{n-1} + H_l \vpsi_{n+1}  = E \vpsi_n,\quad n\in \mathbb{Z}.
\end{gather}
Discrete translational invariance and the Bloch theorem imply the eigenvectors of the equations are plain waves and we get after the Fourier transform:
\begin{gather}
    \mh_k u_k = E_k u_k,\\
    \label{eq:H_k}
    \mh_k = H_0 + H_l e^{-ik} + H_r e^{ik}.
\end{gather}
Diagonalization of this Hamiltonian, analysis of its band structure and detection of various flatbands is the main content of the following sections.

\section{The generator}
\label{sec:gen}

The Hermitian flatband generator with $H_l = H_r^\dagger=H_1$, was introduced and analyzed in Refs.~\onlinecite{maimaiti2017compact,maimaiti2019universal}. It is relies on the compact localized states (CLS) - eigenstates of flatbands with short-range hopping with strictly compact support/occupying a finite number of unit cells - as the main building block. It can be employed straightforwardly in some but not all non-Hermitian cases as we will show below. An alternative method is useful when the spectrum of the Hamiltonian~\eqref{eq:H_k} is known analytically. One then requires that one or more eigenvalues of $\mh_k$~\eqref{eq:H_k} to be $k$-independent. Solving the resulting equations on the matrix elements one recovers the hopping matrices $H_0,H_l, H_r$. This method is particularly useful in the case of small number of bands, practically $\nu<5$, where the entire spectrum can be computed analytically, at least in principle.

\begin{figure}
	\centering
	\includegraphics[width=0.9\columnwidth]{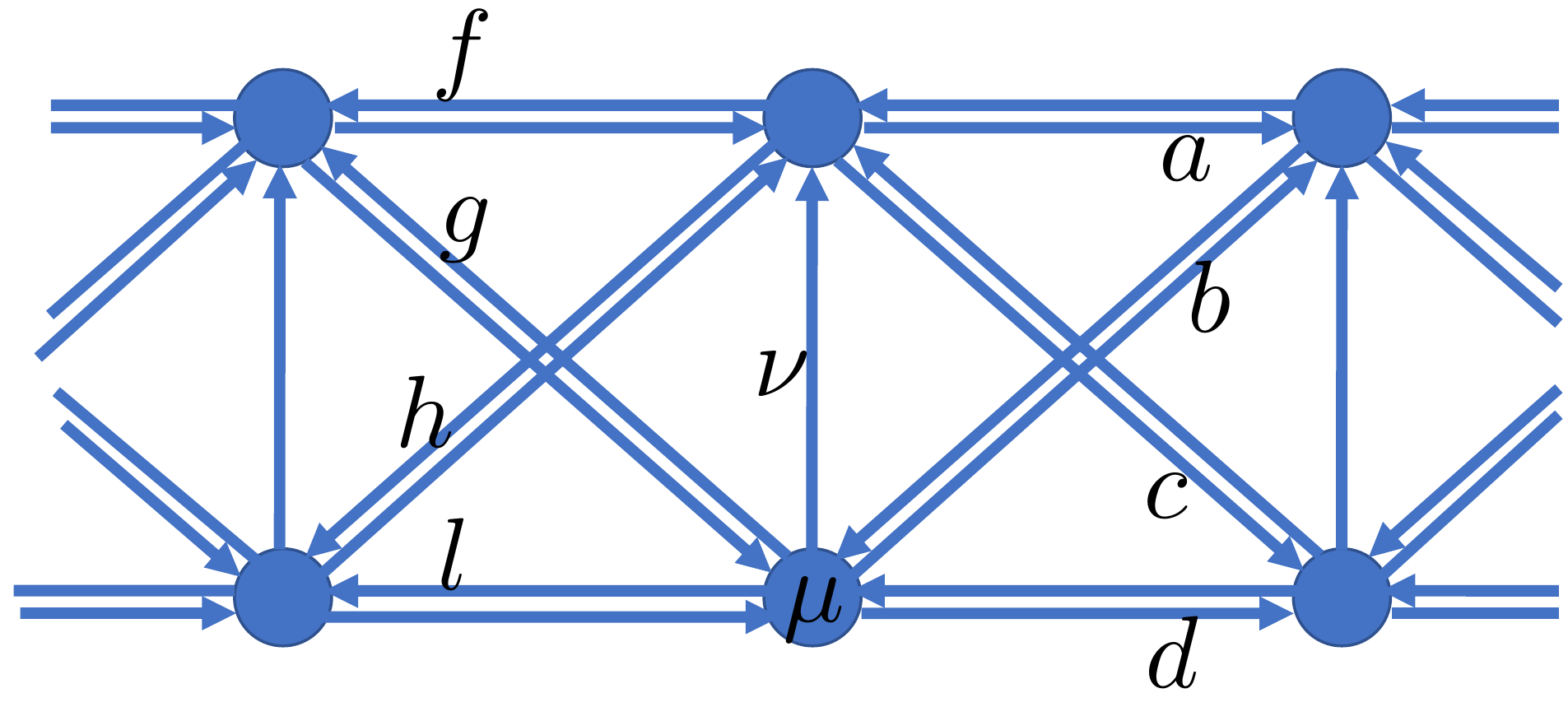}
	\caption{Hoppings of the non-Hermitian lattice~\eqref{eq:eig-nh}: The non-Hermiticity comes form the gain and loss in the hoppings, which results in the asymmetry of the hoppings represented by the arrows.}
	\label{fig:nh-lattice}
\end{figure}

Next we discuss the convenient parameterizations of the hopping matrices $H_{\alpha=0,l,r}$. In the Hermitian case~\cite{maimaiti2017compact} the matrix $H_0$ is always Hermitian and can be taken diagonal in general, because Hamiltonians are equivalent under unitary transformations. In the non-Hermitian case that is no longer true: $H_0$ is not Hermitian in general and might not be diagonalizable. However we can parameterize $H_0$ through the Jordan normal form~\cite{shilov1977linear}
\begin{gather}
    \label{eq:def-H0}
    H_0(\nu,\mu) = \begin{pmatrix}
        0 & \nu\\
        0 & \mu
    \end{pmatrix}\;,
\end{gather}
where $\mu,\nu\in\{0,1\}$ and $\mu\nu=0$, i.e. they cannot be equal to $1$ simultaneously. Here we have used the freedom to shift and rescale the spectrum. This corresponds to three distinct forms of $H_0$:
\begin{align}
    H_0 & = \begin{pmatrix}
        0 & 0\\
        0 & 0
    \end{pmatrix},\quad\mu=\nu=0 - \text{degenerate},\\
    H_0 & = \begin{pmatrix}
        0 & 0\\
        0 & 1
    \end{pmatrix},\quad\nu=0, \mu=1 - \text{non-degenerate,} \label{eq:non-degen-h0}\\
    H_0 & = \begin{pmatrix}
        0 & 1\\
        0 & 0
    \end{pmatrix},\quad\nu=1, \mu=0 - \text{abnormal}.
    \label{eq:def-H0-froms}
\end{align}
The first two cases correspond to diagonalizable $H_0$, while the third one corresponds to non-diagonalizable $H_0$ and is special to the non-Hermitian case. We use the following parameterization for the intercell hopping matrices
\begin{gather}
    H_l = \begin{pmatrix}
        f & g\\
        h & l
    \end{pmatrix},\quad 
    H_r = \begin{pmatrix}
        a & b\\
    c & d
    \end{pmatrix}.
    \label{eq:def-Hlr}
\end{gather} 
Graphical illustration of the non-Hermitian lattice corresponding to above hopping matrices is shown in Fig.~\ref{fig:nh-lattice}. The non-Hermiticity is represented by unidirectional hopping links inside the unit cell. In some cases a suitable similarity transformation might convert these to onsite terms, e.g. gain and loss.

The $k$-space Hamiltonian~\eqref{eq:H_k} is a $2\times 2$ matrix whose eigenvalues $x_k,y_k$ can be computed explicitly. It is more convenient for our purpose to study the following equations
\begin{equation}
\begin{aligned}
    \label{eq:band-eqn}
    x_k + y_k & = \mu + e^{ik}(a + d) + e^{-ik}(f + l)\\
    x_k y_k & = e^{2ik}\det H_r + e^{-2ik}\det H_l \\ 
    & + (\nu f - \mu h)e^{ik} + (\nu a - \mu c)e^{-ik} \\
    & +df - cg - bh + al,
\end{aligned}
\end{equation}
which follow from the Vieta's formulae for the characteristic equation of~\eqref{eq:H_k}.

Our goal is to analyze the solutions of these two equations with respect to the coefficients of $H_r,H_l$ ($a,b,c,d\dots$)  under the suitable constraints on $k$ dependence of either $x_k$ or $y_k$ or both. At this point it is important to note that the non-Hermitian case is richer than the Hermitian one. In the Hermitian case we could only require either of $x_k,y_k$ or both to be $k$-independent to generate \emph{perfect flatband(s)}. For the non-Hermitian system, where $x_k,y_k$ are not necessarily real there are additional options of imposing only real, or imaginary parts of $x_k,y_k$ to be $k$-independent resulting in what we coin as \emph{partial flatbands}. The third option is to request their moduli to be $k$-independent. 

\section{Results}
\label{sec:res}

In this section we discuss one by one the solutions of the system~\eqref{eq:band-eqn} under the constraints on $x_k,y_k$ discussed above. Since full solutions for all the cases take a lot of space, we only discuss specific cases, referring the reader to Appendices for the full details.

\subsection{Perfect flatbands}
\label{sec:pfb}

For perfect flatbands we impose $k$-independence of $x_k,y_k$. This case is the direct analogue of the Hermitian models and we  find compact localized states for the eigenstates. Since these flatbands feature CLS, we adopt the language of the Hermitian flatbands: we refer to a CLS that occupies $U$ unit cells as the class $U$ CLS, and to the respective FB as FB of class $U$.

\subsubsection{Two flatbands}

\begin{figure}
	\centering
	\includegraphics[width=0.9\columnwidth]{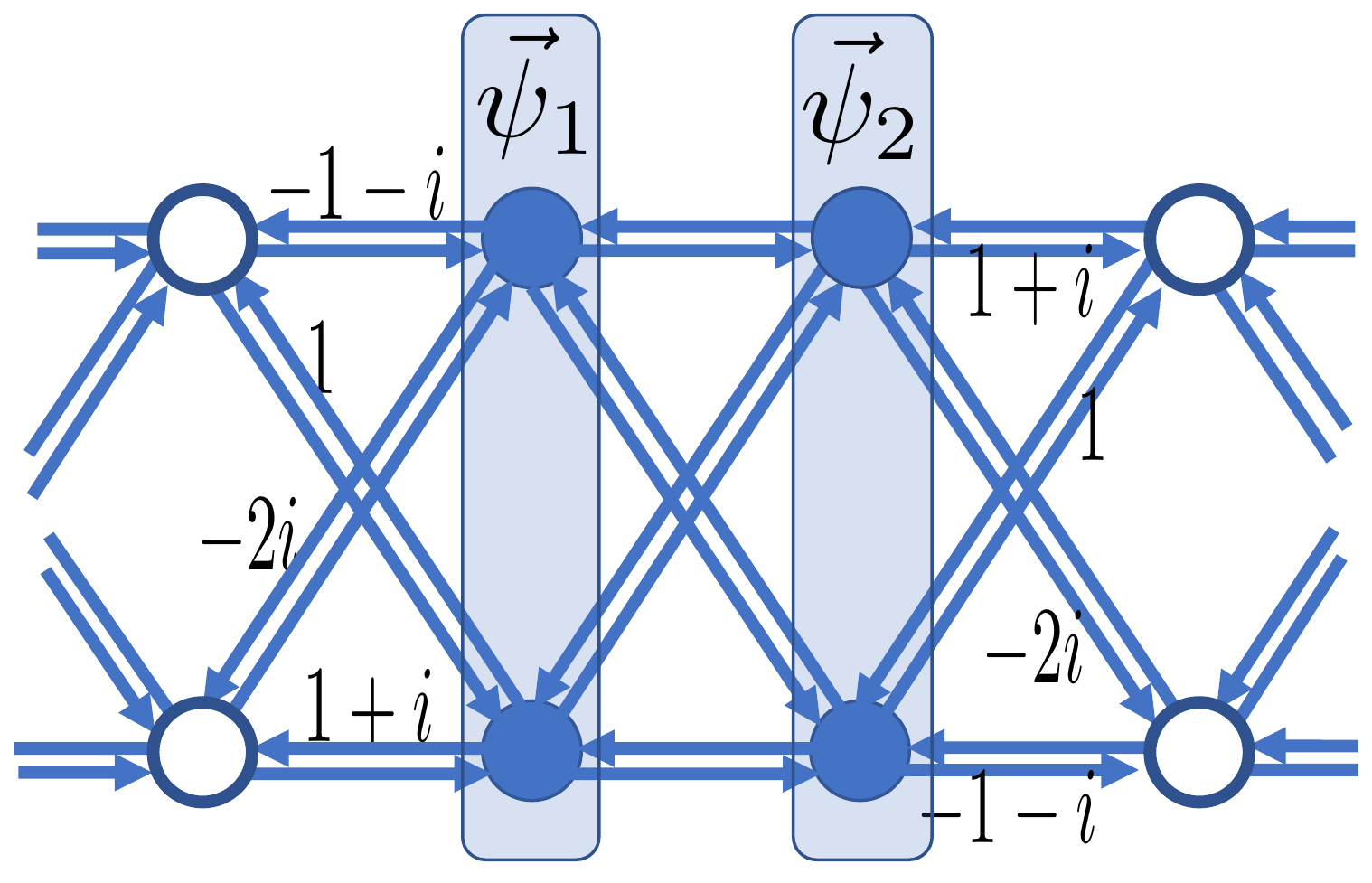}
	\caption{Non-Hermitian lattice with all bands flat: the hopping strengths are as shown next to the hopping links. $\efb=\mp 2 (1-i)$ and CLS size is $U=2$, filled circles indicate the sites of the CLS with non-zero amplitudes - $\vpsi_1=(-1,-1-i), \vpsi_2=\pm(i,1-i)$, and empty circles indicate all the other sites with zero wave amplitude.}
	\label{fig:nh-all-fb-eg}
\end{figure}

In this case we enforce $x_k=x,y_k=y$ in Eq.~\eqref{eq:band-eqn}. Then requiring all the $k$-dependent terms to vanish we get (For the technical details of the derivation see Appendix~\ref{app:both-fb}):
\begin{equation}
\begin{aligned}
    & a + d = 0,\quad f + l = 0,\\
    & \det\,H_r = ad - bc = 0,\\
    & \det\,H_l = fl - hg = 0,\\
    & \nu f - \mu h = 0,\quad \nu a - \mu c = 0,\\
    & xy = df - cg - bh + al,\\
    & x + y= \mu.
    \label{eq:bands-eqn-allf-xy-eqs}
\end{aligned}
\end{equation}
From these it follows $d=-a$, $l=-f$ and either $f=a=0$ (abnormal), or $h=c=0$ (non-degenerate), or $a,c,f,h\neq0$ (degenerate). Also the hopping matrices $H_{l,r}$ are singular just like in the Hermitian case. Solving Eqs.~\eqref{eq:bands-eqn-allf-xy-eqs} for different choices of $H_0$ (Eqs.~\eqref{eq:def-H0-froms}), we can recover $H_l,H_r$ that give two flatbands with energies $\efb = \pm x$ (for details see Appendix~\ref{app:both-fb}). As an illustration we provide the solution for the degenerate $\mu=\nu=0$ case (see Appendix~\ref{app:both-fb} for all the other cases):
\begin{equation}
\begin{aligned}
	\label{eq:nh-all-fb-sol}
	H_r &= \begin{pmatrix}
		a & b\\
		-\frac{a^{2}}{b} & -a
	\end{pmatrix},\\ 
    H_l &= \begin{pmatrix}
        f & g\\
        -\frac{f^{2}}{g} & -f
    \end{pmatrix},\\
    \efb^2 &= 2af - \frac{a^2 g}{b} - \frac{b f^2}{g}\;,
\end{aligned}
\end{equation} 
There are four free parameters $a,b,f,g$ that are in general complex. A specific example is shown in Fig.~\ref{fig:nh-all-fb-eg} with $a = 1+i, b=1, f=-1-i, g=1$, for which we obtain a lattice with two flatbands $\efb=\pm 2(1-i)$ and their respective CLS shown in the figure. Using the CLS tester introduced in Ref.~\onlinecite{maimaiti2017compact}, we find that these CLS have size $U=2$ in general. Unlike the Hermitian case~\cite{maimaiti2017compact,danieli2020cagingI} it does not seem to be possible to detangle this model into non-interacting sites for generic values of $a,b,f,g$.

Previous studies considering non-Hermitian flatbands~\cite{ramezani2017non,leykam2017flat} relied on the \pts  to ensure real eigenenergies. Importantly, the above non-Hermitian Hamiltonian~\eqref{eq:nh-all-fb-sol} can be fine-tuned to have two real flatbands in the absence of \pts. One possible way is to take $a,b,f\in\mathbb{R}$ and set $g=-b$ in Eq.~\eqref{eq:nh-all-fb-sol}. Then the Hamiltonian~\eqref{eq:H_k} becomes 
\begin{gather}
	H_k = \begin{pmatrix}
 		e^{-i k} \left(e^{2 i k} a+f\right) & b e^{-i k} \left(-1+e^{2 i k}\right) \\
 		\frac{e^{-i k} \left(f^2-a^2 e^{2 i k}\right)}{b} & -e^{-i k} \left(e^{2 i k} a+f\right) \\
	\end{pmatrix}
	\label{eq:nh-all-fb-ham}
\end{gather} 
and it has two real flatbands $\efb = \pm (a + f)$. Yet this Hamiltonian is neither Hermitian, nor \pt-symmetric. 

\subsubsection{Single flatband} 

\begin{figure}
	\centering
	\includegraphics[width=0.48\columnwidth]{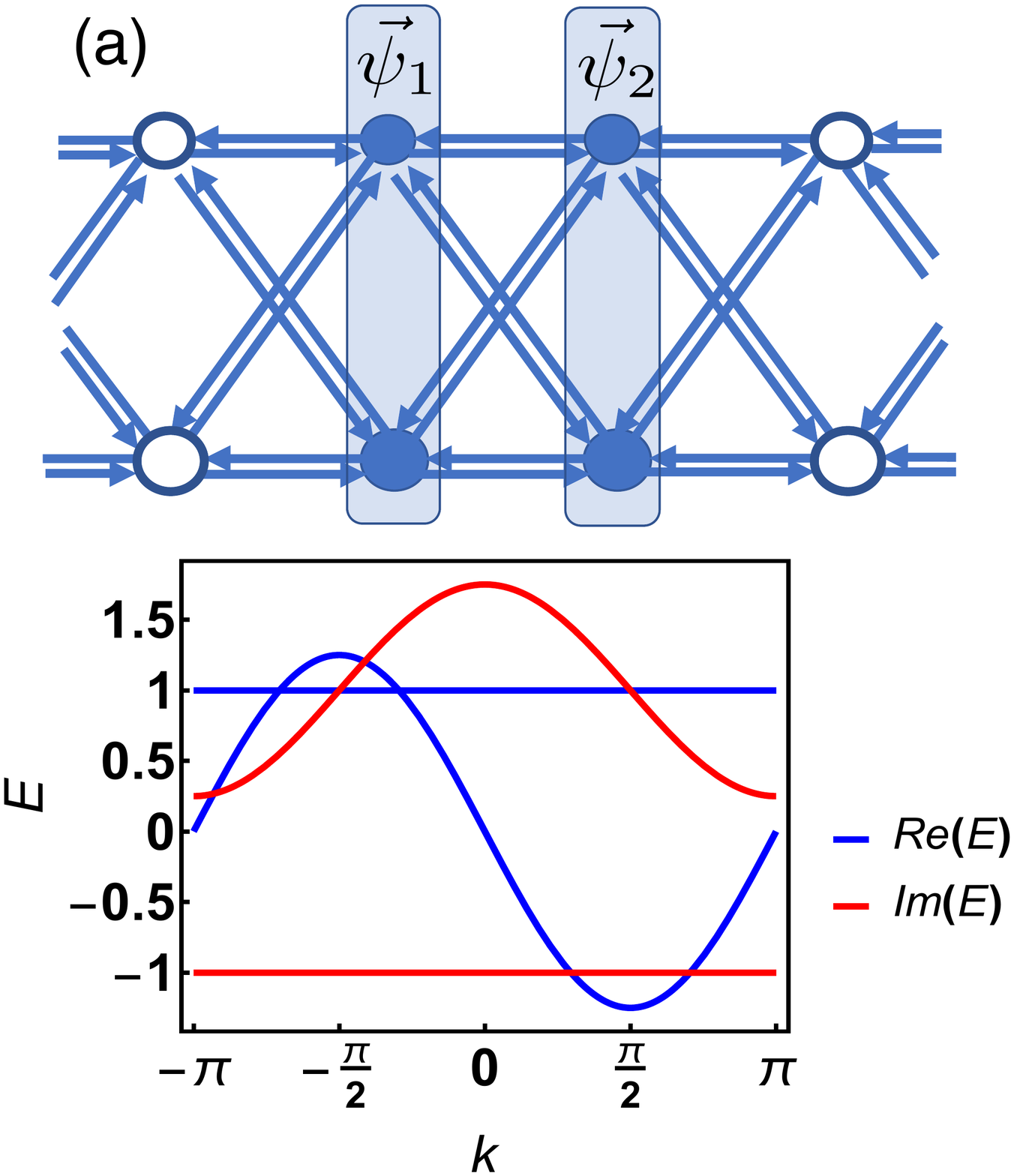}
	\includegraphics[width=0.47\columnwidth]{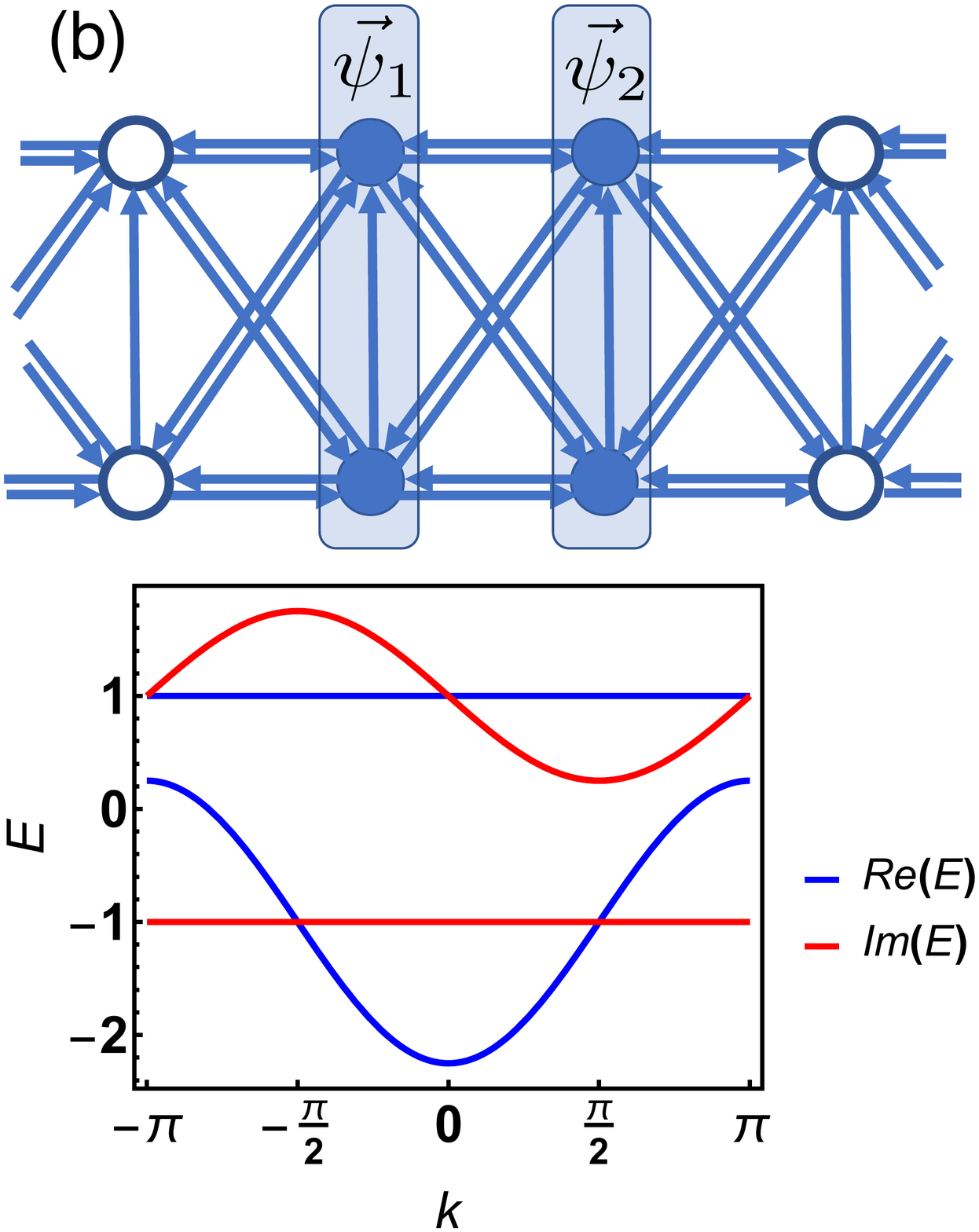}
	\caption{Single flatband of the Hamitlonian~\eqref{eq:Hr-Hl-diff-para}: (a) non-degenerate (canonical) case with parameters(Eq.~\eqref{eq:single-flat-gen-sol}) $a=1+i, x=1-i, f=-\frac{1}{4}-\frac{i}{4}$, (b) abnormal case with parameters $a=1+i, x=1-i, f=-\frac{1}{4}-\frac{i}{4}, g=\frac{1}{4}$. The CLS size is $U=2$ in both cases. Filled circles indicate the sites with non-zero CLS amplitudes, while the empty circles are the sites with zero amplitudes. The amplitudes are given by Eq.~\eqref{eq:u2-cls-app} in Appendix~\ref{app:csl-gen-u2}.}
	\label{fig:nh-single-fb-eg}
\end{figure}

Requiring any of $x$ or $y$ in Eq.~\eqref{eq:band-eqn} to be $k$-independent yields again singular $H_{l,r}$: $\det H_r =0, \det H_l = 0$. Therefore, just as in the case of two flatbands, we can parameterize $H_r,H_l$ as
\begin{gather}
	H_r = \begin{pmatrix}
    		a & b\\
        c & \frac{b c}{a}
    \end{pmatrix}, 
    H_r = \begin{pmatrix}
    		f & g\\
        h & \frac{g h}{f}        
   	\end{pmatrix}, 
    \label{eq:Hr-Hl-diff-para}
\end{gather} 
Substituting $d=\frac{b c}{a}$ in Eq.~\eqref{eq:band-eqn} and requiring $x_k=\efb$ to be $k$-independent gives the following solution (see Appendix~\ref{app:one-band-fb} for details)
\begin{equation}
\begin{aligned}
	b & =\frac{\left(-\efb\pm \sqrt{\efb^2-4 a f}\right) (f \nu +g \efb)+2 a f g}{2 f^2}\\ 
 	c &=\frac{(\efb - \mu) \left(\left(\efb^2\pm \efb \sqrt{\efb^2-4 a f}\right)-2 a f\right)}{2 (f \nu + g \efb)}\\
  	h & =\frac{f^2 (\mu - \efb)}{f \nu + g \efb}.
    \label{eq:single-flat-gen-sol}
\end{aligned}
\end{equation}
The dispersive band is (note the two solutions corresponding to the $\pm$ signs)
\begin{equation}
\begin{aligned}
    E_k &=\mu - \efb +\frac{\left(a e^{i k}+f e^{-i k}\right) (f \nu +g \mu )}{f \nu +g \efb } \\
& -\frac{e^{i k} \nu  (\efb -\mu ) \left(\efb \pm \sqrt{\efb^2-4 a f}\right)}{2 (f \nu +g \efb )}.
	\label{eq:single-flat-gen-band}
\end{aligned}
\end{equation}
Substituting all the possible pairs of values of $\mu,\nu$ into Eqs.~(\ref{eq:single-flat-gen-sol}-\ref{eq:single-flat-gen-band}) we generate single flatband Hamiltonians for the degenerate, non-degenerate and abnormal cases and their respective band structures, as illustrated in Fig.~\ref{fig:nh-single-fb-eg}. Interestingly the degenerate case, $\mu=\nu=0$ reduces to the previously considered case of two flatbands, with $E=\pm x$. There are $4$ free complex parameters: $x,a,f,g$ while in the Hermitian case there are only two real free parameters.~\cite{maimaiti2017compact} The CLS corresponding to this FB are in general of size $U=2$ (see also Sec.~\ref{sec:cls-gen} and Appendix~\ref{app:csl-gen-u2}). Also unlike the Hermitian case where a FB in a two band system is always gaped away from the dispersive band, in the non-Hermitian setting, the FB and the dispersive bands can have crossings of the real and imaginary parts of the energy as illustrated in Fig.~\ref{fig:nh-single-fb-eg}(a). 

\subsection{CLS based construction}
\label{sec:cls-gen}

We also rederive the above results with the help of the CLS based method originally developed for the Hermitian models~\cite{maimaiti2017compact,maimaiti2019universal}. We sketch briefly the straightforward modification of the construction for the Hermitian case and refer the interested reader to Appendix~\ref{app:cls-gen} which contains the full details of the method. We search for a non-Hermitian Hamiltonian with the CLS $\PCLS=(\vpsi_1,\vpsi_2,\dots,\vpsi_U)$ of size $U$. This CLS is an eigenvector of the $U \times U$ block tridiagonal matrix
\begin{gather}
    \mh_U = \begin{pmatrix}
        H_0 & H_l & 0 & 0 & \dots & 0\\
        H_r & H_0 & H_l & 0 & \dots & 0\\
        0 & \ddots & \ddots & \ddots & \ddots & \vdots\\
        \vdots & ~ & ~ & ~ & ~ & \vdots\\
        0 & \dots & 0 & H_r & H_0 & H_l\\
        0 & \dots & 0 & 0 & H_r & H_0
    \end{pmatrix}\; 
    \label{eq:cls-def-HU}
\end{gather}
with eigenenergy $\efb\in\mathbb{C}$. Out of the $U$ eigenvectors of this Hamiltonians the CLS is selected by the destructive interference conditions
\begin{gather}
    H_l\kpsi{1} = H_r\kpsi{U} = 0,
    \label{eq:dest-intf-cond}
\end{gather}
that ensure that the eigenstate remains compactly localized under the action of $\mh_U$. Therefore a \emph{necessary} condition for existence of a non-Hermitian CLS reads
\begin{gather}
    \det H_l = \det H_r = 0.
    \label{eq:necessary-cond-cls}
\end{gather} 
As in the Hermitian case, we rewrite the CLS eigenproblem~(\ref{eq:cls-def-HU}-\ref{eq:dest-intf-cond}) as 
\begin{align}
    \label{eig-1}
    H_l\kpsi{2} &= (\efb - H_0)\kpsi{1}\\
    \label{eig-2}
    H_r\kpsi{j-1} + H_l\kpsi{j+1} &= (\efb - H_0)\kpsi{j}, 2 \le j\le U-1\\
   \label{eig-3} 
    H_r\kpsi{U-1} &= (\efb - H_0)\kpsi{U}\\
    \label{eig-4}
    H_l\kpsi{1} &= H_r\kpsi{U} = 0\\
    \label{eig-5}
    \kpsi{j} &= 0 \;,\;  j<0,\,j>U \;.
\end{align}
The non-Hermitian FB generator is the set of all possible matrices $H_0,H_r,H_l$ and CLSs $\PCLS=(\vpsi_1,\vpsi_2,\dots,\vpsi_U)$ that satisfy the equations (\ref{eig-1}-\ref{eig-5}). It is solved with the same inverse eigenvalue problem as in the Hermitian case,~\cite{maimaiti2019universal} e.g. treating $\vpsi_{i=1\dots U}$ and $\efb$ as inputs and reconstructing $H_{l,r}$. It is worth pointing out that $H_0$ does not have to be taken diagonal in this setting. This formulation is especially convenient for constructing the most common type of non-Hermitian type of Hamiltonians, where $H_l=H_r^\dagger$, e.g. the intercell hopping is Hermitian, but the onsite terms are not, i.e. diagonal of $H_0$ contains gain and loss terms. 

As an example we provide solution for the $U=2$ case. The solution of this eigenproblem in the case of two bands $\nu=2$ and $H_0$ given by Eq.~\eqref{eq:def-H0} is presented in Appendix~\ref{app:cls-gen}.

In the $U=2$ case the eigensystem~\eqref{eig-1} simplifies to
\begin{equation}
\begin{aligned}
    & H_l\kpsi{2} = \left(\efb - H_0\right)\kpsi{1}\\
    & H_r\kpsi{1} = \left(\efb - H_0\right)\kpsi{2}\\
    & H_l\kpsi{1} = H_r\kpsi{2} = 0.
\end{aligned}
\end{equation}
Unlike the Hermitian case,~\cite{maimaiti2019universal} now the $U=2$ case decouples into two independent inverse eigenvalue problems for $H_l$ and $H_r$. Their solutions are
\begin{equation}
    \begin{aligned}
        & H_l = \frac{\left(\efb - H_0\right)\kpsi{1}\bpsi{2}Q_1}{\mel{\psi_2}{Q_1}{\psi_2}}\\
        & H_r = \frac{\left(\efb - H_0\right)\kpsi{2}\bpsi{1}Q_2}{\mel{\psi_1}{Q_2}{\psi_1}}
    \end{aligned}
\end{equation}
Similarly to the Hermitian case~\cite{maimaiti2017compact}, the maximum size of a CLS is $U=2$. This is confirmed by the consistency of the $U=2$ solution with the solution from the band calculation method that we presented above and which contains all possible two band non-Hermitian FB Hamiltonians.

\subsection{Partial flatbands}
\label{sec:pfb}

The case of partial flatbands when only real, imaginary parts or modulus of the eigenenergy are $k$-independent is only possible for non-Hermitian Hamiltonians. Unlike the case of perfect flatbands the eigenstates of partial flatbands are not compactly localized, since the bands are not fully flat. Therefore the CLS construction cannot be employed in this case and we use the band calculation method, which also restricts us to the case of $2$ bands only.

We write $x_k=x_1 + i x_2, y=y_1 + i y_2$, where $x_i,y_i\in\mathbb{R},\,i=1,2$ and write the real and imaginary parts of Eq.~\eqref{eq:band-eqn} separately
\begin{equation}
\begin{aligned}
	\label{eq:partial-flat-gen}
    x_1 + y_1 &= \mu + (a + d + f + l)\cos k\\
    x_2 + y_2 &= -(a + d - f - l)\sin k\\
    x_1 y_1 - x_2 y_2 &= al - bh - cg + df\\
    & +(a\mu - c\nu + f\mu - h\nu)\cos k\\ 
    & +(\det H_l + \det H_r)\cos 2k\\
    x_2 y_1 + x_1 y_2 &= \left(-a\mu + c\nu + f\mu - h\nu\right)\sin k\\ 
    & +(\det H_l -\det H_r)\sin 2k\;
\end{aligned}
\end{equation}
Requiring different subsets of $x_1,x_2,y_1,y_2$ to be $k$-independent and solving the above set of equations with respect to the hopping matrices $H_r,H_r$ we can find Hmiltonians with partial flatbands. There are many possible subsets for partial flatbands: real(imaginary) parts of both bands are flat, real(imaginary) part of only one band is flat, real(imaginary) part of one band and imaginary(real) part of the other band are flat, etc. For convenience we only consider two specific cases of partial flatbands, while the full details are provided in the Appendix~\ref{app:partial-fb}. 

\paragraph{Real parts of two bands are flat} 

\begin{figure}[htb!]
	\centering
	\includegraphics[width=0.8\columnwidth]{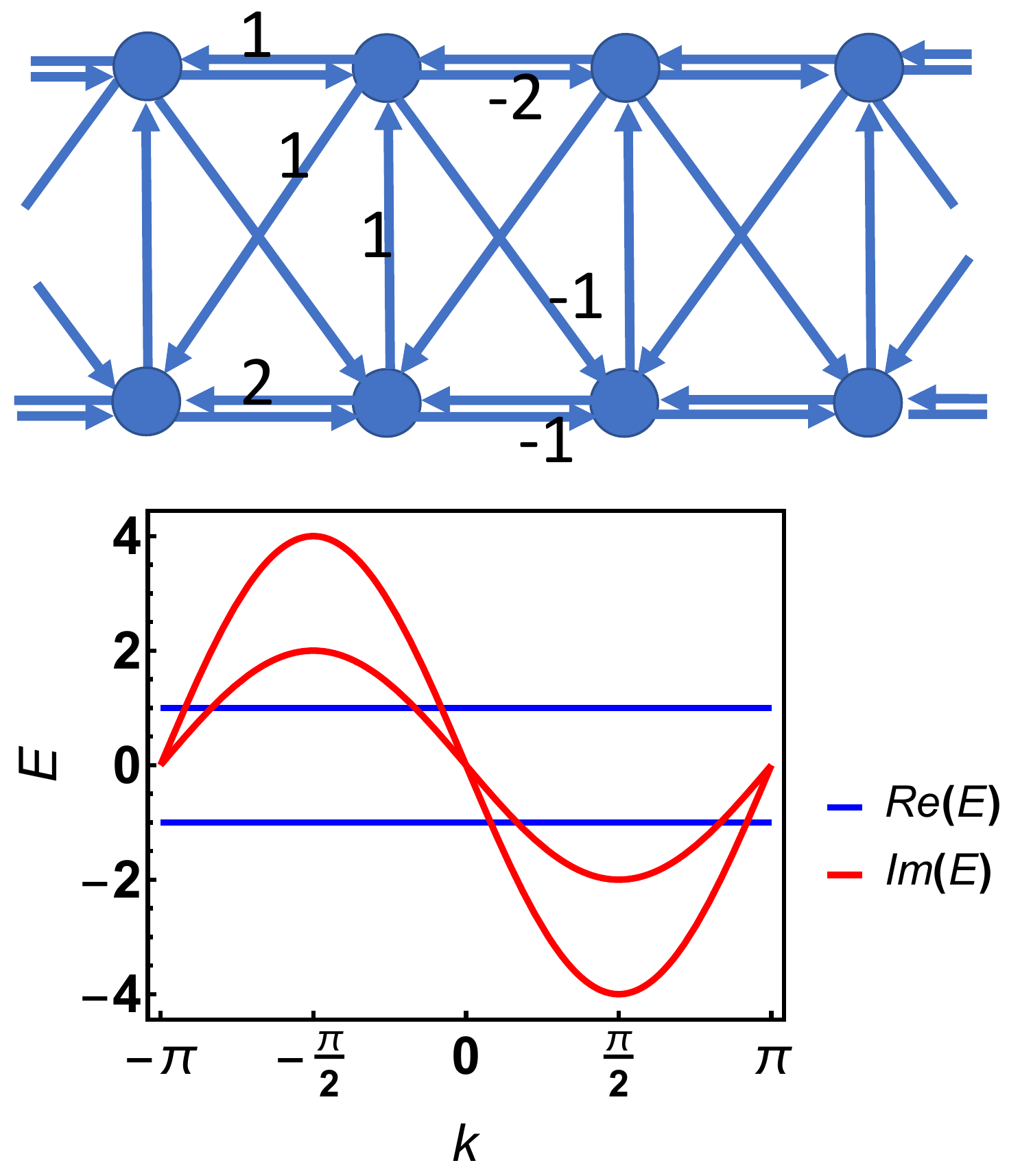}
	\caption{Example of real parts of both bands flat case with abnormal $H_0$. The parameters in these examples are $x_1= 1,d= -1,f= 1,h= 1$. Corresponding hopping strengths are also given in the figure.}
	\label{fig:all-re-fb-eg}
\end{figure}

In this case $x_1,y_1$ in Eq.~\eqref{eq:partial-flat-gen} are $k$-independent. We consider here the abnormal $H_0$, i.e. $\nu=1,\mu=0$. Solving the system~\eqref{eq:partial-flat-gen} we find the following solution (see Appendix~\ref{app:both-re-fb} for further details):
\begin{equation}
\begin{aligned}
    H_{0}&=\begin{pmatrix}0 & 1\\
                        0 & 0
                \end{pmatrix},\\
         H_{l}&=\begin{pmatrix} f & \frac{h^2-x_1^2 \left(d+f-x_1\right){}^2}{4 h x_1^2}\\
                        h & x_1-d
               \end{pmatrix},\\
         H_{r}&=\begin{pmatrix}-f-x_1 & \frac{x_1^2 \left(d+f+x_1\right){}^2-h^2}{4 h x_1^2}\\
                        -h & d
               \end{pmatrix}\; ,
	\label{eq:both-real-flat-abnormal}
\end{aligned}
\end{equation}
The corresponding band structure is
\begin{equation}
\begin{aligned}
	E_1 & = -x_1+i \sin (k) \left(d-f+\frac{h}{x_1}-x_1\right)\\
        E_2 & = x_1-i \sin (k) \frac{ \left(x_1 \left(-d+f+x_1\right)+h\right)}{x_1} \; ,
    \label{eq:both-rl-fb-abn-band}
\end{aligned} 
\end{equation}
where the parameters $x_1, d, f, h$ are real. Figure~\ref{fig:all-re-fb-eg} shows a specific example with $x_1= 1,d= -1,f= 1,h= 1$. Because of the flat real part of the eigenenergy, it might seem that kinetic energy is quenched and no transport is possible in that band. However, one might observe an apparent spreading of an initially localized state due to the $k$-dependence of the imaginary part of the eigenenergy. If the original state overlaps with several eigenmodes, the mode with the largest imaginary part would be amplified (or decay the slowest if all the eigenmodes are decaying). As the eigenmodes are extended this might lead to a spreading of the initial state.

\paragraph{Modulus of a band is flat}

In this case, we assume that one of the bands has constant modulus, but allow the phase to be $k$-dependent: $x=r e^{i\theta_k}, r\in \mathbb{R}$ in Eq.~\eqref{eq:band-eqn}. Since equation~\eqref{eq:band-eqn} only involves integer powers of $e^{i k}$, this implies that the phase is also an integer multiple of $k$, i.e. $\theta_k=m k$, and $m$ can only take two  values $m=\pm 1$ (see Appendix~\ref{app:mod-fb}). For each value of $m$ we solve Eq.~\eqref{eq:band-eqn} and get the hopping matrices $H_r,H_l$ that gives the band structure $E= r e^{i\theta_k}, r\in \mathbb{R}$. As an example we present results for $m=1$ and abnormal $H_0$, i.e. $\mu=0,\nu=1$:
\begin{equation}
\begin{aligned}
    H_r & = \begin{pmatrix}
        r & b\\
        0 & d
    \end{pmatrix}\\
    H_l & = \begin{pmatrix}
        0 & g\\
        0 & l
    \end{pmatrix},
    \label{eq:mod-fl-result}
\end{aligned}
\end{equation}
which has the following band structure
\begin{align}
    \label{eq:mod-fl-band}
    E_1 & = e^{ik}r,\\
    E_2 & = de^{ik} + e^{-ik}l + 1.\notag
\end{align}  
Similarly, other cases of $H_0$ and $m=-1$ case are solved (see details in Appendix~\ref{app:mod-fb}). Interestingly the above band $E_1$ has a non-trivial winding number which is expected since this only part of the band energy that can change. 

\section{Conclusions}
\label{sec:conclusion}

We presented a systematic way of constructing flatbands in non-Hermitian networks with two or more bands. The existence of these flatbands relies on fine-tuning of the hoppings and/or onsite energies rather than some specific symmetry, like chiral or parity-time symmetries. We considered two distinct cases of perfect and partial flatbands. The former just like their Hermitian counterparts host compact localized states that can be used to construct these bands. We focused in this work on the   case of two band networks and we discovered that the maximum possible size of the CLS is $U=2$ for the perfect flatbands, similarly as in the Hermitian case. We have also provided examples of real valued perfect flatbands in non-Hermitian models not related to any symmetry. The other case considered was partial flatbands that are absent in the Hermitian models and are special to the non-Hermitian setting with either real or imaginary part or modulus of the eigenenergies becoming dispersionless. Better understanding of their physical properties is an interesting open problem. Interestingly, in the case of constant modulus of the eigenenergy we observed that the respective band has a non-trivial winding number. 

Our work is only an initial step in systematic understanding non-Hermitian flatbands. The obvious questions are the connection to Hermitian flatbands - how could one perturb them systematically with gain/loss terms and keep the flatband? What is the effect of various perturbations -- disorder or interaction -- on the non-Hermitian flatbands? The partial flatbands do not have analogies in Hermitian models and are therefore of special interest. The case of flat imaginary part of eigenergies (see appendix~\ref{app:single-im-fb}) might be interesting, as it might exhibit dynamics that is sensitive to initial conditions, due to the presence of gain/loss and in contrast to the real part of eigenenergy flat case. It might be interesting to study this case further, to see whether the propagation dynamics remain qualitatively similar to the Hermitian case, or if new features emerge due to non-Hermiticity. In cases of real perfect flatbands we might expect appearance in finite chains of edge states with complex energies, or more dramatically, non-Hermitian skin effect, in which the bulk eigenvalues are highly sensitive to the boundary conditions.


\begin{acknowledgments}
    We are grateful to Sergej Flach and Daniel Leykam for their valuable feedback on the manuscript. This work was supported by the Institute for Basic Science in Korea (IBS-R024-D1).
\end{acknowledgments}

\appendix 

\section{CLS based generator} 
\label{app:cls-gen}

In this appendix we show the CLS based derivation of the perfect flatbands.

\subsection{The $U=1$ case}

In this case the eigenvalue problem (\ref{eig-1}-\ref{eig-5}) becomes
\begin{equation}
\begin{aligned}
    & H_0 \vpsi = \efb\vpsi\\
    & H_r \vpsi = H_l \vpsi = 0
    \label{eq:cls-eq-u1}
\end{aligned}
\end{equation} 
Parameterizing $\vpsi = (x, y) \ne 0$, Eq.~\eqref{eq:cls-eq-u1} becomes
\begin{equation}
\begin{aligned}
    \nu y & = \efb x \\
    \mu y & = \efb y \\
    a x + b y & = \efb x \\
    c x + d y & = \efb y \\
    f x + g y & = \efb x \\
    h x + l y & = \efb y \; .
    \label{eq:cls-eq-u1-1}
\end{aligned}
\end{equation}
Next we solve this system for all the possible choices of $H_0$. 

\paragraph{Degenerate $H_0$} 

In this case $\mu=\nu=0$ and Eq.~\eqref{eq:cls-eq-u1-1} yields the following solution
\begin{align*}
    & \efb = 0\\
    & a  = -\frac{y}{x} b, \quad c  = -\frac{y}{x} d\\
    & f  =-\frac{y}{x} g, \quad h  =-\frac{y}{x} l\\
\end{align*}
Therefore the hopping matrices read
\begin{gather*}
    H_0 = \begin{pmatrix}
        0 & 0\\
        0 & 0
    \end{pmatrix},\\
    H_l = \begin{pmatrix}
        -\frac{y}{x} g & g\\
        -\frac{y}{x} l & l
    \end{pmatrix},\quad
    H_r = \begin{pmatrix}
        -\frac{y}{x} b & b\\
        -\frac{y}{x} d & d
    \end{pmatrix},\\
\end{gather*}

\paragraph{Abnormal $H_0$} 

In this case $\mu=0,\nu=1$ and Eq.~\eqref{eq:cls-eq-u1-1} gives the following solution $y = a = c = f = h = 0$.
Therefore the hopping matrices, CLS, and FB energy are
\begin{gather*}
    H_0 = \begin{pmatrix}
        0 & 1\\
        0 & 0
    \end{pmatrix},\\
    H_l = \begin{pmatrix}
        0 & g\\
        0 & l
    \end{pmatrix},\quad
    H_r = \begin{pmatrix}
        0 & b\\
        0 & d
    \end{pmatrix},\\
    \vpsi = (x, 0),\quad \efb = 0.
\end{gather*} 

\paragraph{Non-degenerate $H_0$}

In this case $\mu=1,\nu=0$ and Eq.~\eqref{eq:cls-eq-u1} have the following solution
\begin{gather*}
    H_0 = \begin{pmatrix}
        0 & 0\\
        0 & 1
    \end{pmatrix},\\
    H_l = \begin{pmatrix}
        f & 0\\
        g & 0
    \end{pmatrix},\quad
    H_r = \begin{pmatrix}
        a & 0\\
        c & 0
    \end{pmatrix},\\
    \vpsi = (0, 1), \quad
    \efb = 1 \; .
\end{gather*} 
Here $y$ has been normalized to $1$.

\subsection{U=2 case}
\label{app:csl-gen-u2}

In this case the eigenvalue problem~(\ref{eig-1}-\ref{eig-5}) becomes
\begin{equation}
\begin{aligned}
    & H_0 \psi_1 + H_l \psi_2 = \efb \psi_1 \\
    & H_0 \psi_2 + H_r \psi_1 = \efb \psi_2 \\
    & H_l \psi_1 = 0,\ \ H_r \psi_2 =0
    \label{u2_eigv_prb}
\end{aligned}
\end{equation}
We can parameterize $H_r, H_l$ as follows
\begin{gather}
    H_r = \begin{pmatrix}
        a & b\\
        c & \frac{b c}{a}
    \end{pmatrix}
    \quad 
    H_l = \begin{pmatrix}
        f & g\\
        h & \frac{g h}{f}
    \end{pmatrix},
    \label{Hr-Hl-dif-para}
\end{gather}
which satisfy destructive interference conditions by definition. Then we can choose $\psi_1, \psi_2$ to be zero eigenvector of $H_l, H_r$ respectively.
\begin{gather}
    \psi_1 = \begin{pmatrix}
        -g\
        f 
    \end{pmatrix},\quad
    \psi_2 = \alpha \begin{pmatrix}
        -b\\
        a
    \end{pmatrix}
    \label{eq:u2-cls-app}
\end{gather} 
Then Eqs.~\eqref{u2_eigv_prb} become
\begin{equation}
\begin{aligned}
    \begin{pmatrix}
        \alpha a  g - \alpha  b f + f \nu\\
        \frac{ \alpha a g h}{f} - \alpha  b h+f \mu
    \end{pmatrix} &=
    \efb \begin{pmatrix}
        -g\\
        f
    \end{pmatrix}\\ 
    \begin{pmatrix}
        \alpha a  \nu -a g+b f\\
        \alpha a  \mu +\frac{b c f}{a}-c g
    \end{pmatrix} &=
    \alpha \efb \begin{pmatrix}
        -b\\
        a
    \end{pmatrix}
    \label{u2_eigv_prob_1}
\end{aligned}
\end{equation} 
Solving Eqs.~\eqref{u2_eigv_prob_1} above we get ($\lambda=\efb$)
\begin{equation}
    \begin{aligned}
        b &= \frac{\left(-\lambda \pm \sqrt{\lambda ^2-4 a f}\right) (f \nu +g \lambda )+2 a f g}{2 f^2}\\
        c &= \frac{(\lambda -\mu ) \left(\left(\lambda ^2\pm \lambda  \sqrt{\lambda ^2-4 a f}\right)-2 a f\right)}{2 (f \nu +g \lambda )}\\
        h &= \frac{f^2 (\mu -\lambda )}{f \nu +g \lambda }\\
        \alpha &= -\frac{\lambda \pm \sqrt{\lambda ^2-4 a f}}{2 a}.
    \end{aligned}
    \label{u2_sol}
\end{equation} 
Putting the corresponding values of $\mu,\nu$ in Eq.~\eqref{u2_sol} we get solutions for degenerate, non-degenerate and abnormal cases. Obviously, the above solution~\eqref{u2_sol} is the same with the solution~\eqref{eq:single-flat-gen-sol} from band calculation method. Since the solution of band calculation method contains all possible $U$ class, $U=2$ is the maximum CLS class. The band structure corresponding to above solution~\eqref{u2_sol} is
\begin{gather*}
    \begin{aligned}
        \efb &= \lambda \\
        E_k &= \mu-\lambda +\frac{\left(a e^{i k}+f e^{-i k}\right) (f \nu +g \mu )}{f \nu +g \lambda } \\
        & -\frac{e^{i k} \nu  (\lambda -\mu ) \left(\lambda \pm \sqrt{\lambda ^2-4 a f}\right)}{2 (f \nu +g \lambda )}.
    \end{aligned}
\end{gather*}

\section{Perfect flatbands: Band calculation method} 

A completely flatband has $k$-independent real and imaginary parts. Our starting point solving this case is equation~\eqref{eq:band-eqn}, which is 
\begin{equation}
\begin{aligned}
    x_k + y_k & = \mu + e^{ik}(a + d) + e^{-ik}(f + l)\\
    x_k y_k & = e^{2ik}\det H_r + e^{-2ik}\det H_l\\
    & + (\nu f - \mu h) e^{ik} + (\nu a - \mu c)e^{-ik}\\
    & + df-cg-bh+al\;.
\end{aligned} 
    \label{eq:band-eqn-app} 
\end{equation}
We assume one of $x_k, y_k$ or both are $k$ independent and solve the above system to find flatband Hamiltonian.

\subsection{The case of both bands are completely flat}
\label{app:both-fb}

In this case both $x_k, y_k$ in Eqs.~\eqref{eq:band-eqn-app} are $k$-independent. We assume $x_k=x$ and $y_k=y$, then Eq.~\eqref{eq:band-eqn-app} becomes
\begin{equation}
\begin{aligned}
    x + y & = \mu + e^{ik}(a+d)+e^{-ik}(f+l)\\
    xy & =e^{2ik}\det H_r + e^{-2ik}\det H_l + (\nu f-\mu h)e^{ik}\\
    & +(\nu a - \mu c)e^{-ik} + df - cg - bh + al.
    \label{eq:bands-eqn-allf-xy}    
\end{aligned}
\end{equation}
Requiring polynomial of $e^{ik}$ to vanish gives the following equations:
\begin{equation}
\begin{aligned}
    & a+d=0,\\
    & f+l=0,\\
    & \det\,H_{r}=ad-bc=0,\\
    & \det\,H_{l}=fl-hg=0,\\
    & \nu f-\mu h=0,\\
    & \nu a-\mu c=0,\\
    & xy=df-cg-bh+al,\\
    & x+y= \mu .
    \label{eq:bands-eqn-allf-xy-eqs-app}
\end{aligned} 
\end{equation}
From these it follows $d=-a$, $l=-f$ and either $f=a=0$, or $h=c=0$, or none of $a,c,f,h$ is zero.

Solving Eq.~\eqref{eq:bands-eqn-allf-xy-eqs-app} for degenerate, non-degenerate and abnormal cases separately, we can get $H_l,H_r$ that gives both bands flat.

\paragraph{Degenerate case}

In this case we have $\mu=\nu=0$ and $bc+a^{2}=0$, $hg+f^{2}=0$, $y=-x$. Therefore $c=-a^{2}/b$, $h=-f^{2}/g$:
\begin{equation}
    \label{eq:bands-eqn-allf-xy-Hlr-sol-zero}
\begin{aligned}
    H_r &= \begin{pmatrix}
        a & b\\
        -\frac{a^{2}}{b} & -a
    \end{pmatrix},\\
    H_l &= \begin{pmatrix}
        f & g\\
        -\frac{f^{2}}{g} & -f
    \end{pmatrix},\\
    x^2 &= 2af-\frac{a^{2}g}{b}-\frac{bf^{2}}{g}.
\end{aligned}
\end{equation} 
Then the corresponding Hamiltonian has two FBs at energies $\efb=\pm x$.

\paragraph{Non-degenerate $H_0$}

In this case $\mu=1,\nu=0$ in Eqs.~\eqref{eq:bands-eqn-allf-xy-eqs-app}. Then Eq.~\eqref{eq:bands-eqn-allf-xy-eqs-app} give $h=c=0$, $ad=fl=0$, and $y=1-x$. Therefore $a=d=f=l=0$ and either $b$ or $c$ are non-zero and either $h$ or $g$ are non-zero:
\begin{equation}
    \label{eq:bands-allf-xy-Hlr-sol-ndeg}
\begin{aligned}
    H_r &= \begin{pmatrix}
        0 & b\\
        0 & 0
    \end{pmatrix}
    \quad\text{or}\quad
    \begin{pmatrix}
        0 & 0\\
        c & 0
    \end{pmatrix},\\
    H_l &= \begin{pmatrix}
        0 & g\\
        0 & 0
    \end{pmatrix}
    \quad\text{or}\quad
    \begin{pmatrix}
        0 & 0\\
        h & 0
    \end{pmatrix},\\
    x(1 - x) &= -cg - bh.
\end{aligned}
\end{equation}
There are four possible solutions of $bc=0,hg=0$. Interestingly, two out of the four, imply $x=0,y=1$ or $x=1,y=0$. Since the zero eigenvectors of $H_l, H_r$ are proportional and they are also an eigenvector of $H_0$ corresponding to eigenvalue $1$ and they satisfy Eq.~\eqref{eq:cls-eq-u1}, the CLS class is $U=1$.

\paragraph{Abnormal $H_0$}

In this case $\mu=0,\nu=1$ and $f=a=0$, and $y=-x$. Therefore $d=l=0$, and $bc=hg=0$, and
\begin{equation}
    \label{eq:bands-eqn-allf-xy-Hlr-sol-an}
\begin{aligned}
    H_r &= \begin{pmatrix}
        0 & b\\
        0 & 0
    \end{pmatrix}
    \quad\text{or}\quad
    \begin{pmatrix}
        0 & 0\\
        c & 0
    \end{pmatrix},\\
    H_l &= \begin{pmatrix}
        0 & g\\
        0 & 0
    \end{pmatrix}
    \quad\text{or}\quad
    \begin{pmatrix}
        0 & 0\\
        h & 0
    \end{pmatrix},\\
    x^2 &= cg + bh.
\end{aligned} 
\end{equation} 
Similarly to the non-degenerate case in previous paragraph, above solution also $U=1$ class.

\subsection{Checking the CLS class of both bands are completely flat case} 
\label{app:all-fb-cls-check} 

The solutions~\eqref{eq:bands-eqn-allf-xy-Hlr-sol-zero} for the degenerate all bands flat case do not tell the CLS class. Here we show a test procedure to check the CLS class of the given solution. 

According to the destructive interference condition~\eqref{eig-5}, the zero eigenvectors of $H_l, H_r$ are $\vpsi_1, \vpsi_U$ respectively. Thus we define
\begin{equation}
	\vpsi_1 = 
	\begin{pmatrix}
	    -g \\ f
	\end{pmatrix},
	\quad
	\vpsi_U = \alpha
	\begin{pmatrix}
	    -b \\ a
	\end{pmatrix}
\end{equation} 
Suppose the CLS class is U=2 then we have $\vpsi_2 = \vpsi_U$. Then the eigenvalue Eq.~\eqref{u2_eigv_prb} are always satisfied with $\alpha=\pm \frac{i \sqrt{g}}{\sqrt{b}}$. Therefore the CLS class is $U=2$.

\subsection{Single perfect flatband}
\label{app:one-band-fb}

Requiring any of $x$ or $y$ in Eq.~\eqref{eq:band-eqn-app} will yield $\det H_r =0, \det H_l = 0$, therefore we can parameterize $H_r,H_l$ as
\begin{equation}
    H_r = \begin{pmatrix}
        a & b\\
        c & \frac{b c}{a}
    \end{pmatrix},
    \quad
    H_r = \begin{pmatrix}
        f & g\\
        h & \frac{g h}{f}
    \end{pmatrix}, 
\end{equation} 
which makes $H_r,H_l$ to be singular by definition. I assume that only $x_{k}=x$ is flat, then Eq.~\eqref{eq:band-eqn-app} becomes
\begin{equation}
    \begin{aligned}
        x+y_k &= \frac{b c e^{i k}}{a}+a e^{i k}+e^{-i k} \left(\frac{g h}{f}+f\right)+\mu)\\
        x y_k &= \frac{(a g-b f) (a h-c f)}{a f} +e^{i k} (a \mu -c \nu )\\ &+e^{-i k} (f \mu -h \nu ).
    \end{aligned}
    \label{eq:bands-eqn-allf-xy-1}
\end{equation}
This results into the following equations
\begin{equation}
    \begin{aligned}
        y_{k} & =\frac{b c e^{i k}}{a}+a e^{i k}+e^{-i k} \left(\frac{g h}{f}+f\right)+\mu -x\\
        y_{k} & =\frac{e^{i k} (a \mu -c \nu )+e^{-i k} (f \mu -h \nu )}{x} \\
        & + \frac{(a g-b f) (a h-c f)}{a f x}  \; .
    \end{aligned}
    \label{eq:bands-eqn-of-generic-2}
\end{equation}
Consequently, equating powers of $e^{ik}$, I find 
\begin{equation}
    \begin{aligned}
        & \frac{b c}{a}+a=\frac{a \mu -c \nu }{x}\\
        & \frac{g h}{f}+f=\frac{f \mu -h \nu }{x}\\
        & \mu -x=\frac{(a g-b f) (a h-c f)}{x (a f)}
    \end{aligned}
    \label{eq:bands-of-generic-3}
\end{equation}

Solving Eq.~\eqref{eq:bands-of-generic-3}
\begin{equation}
    \begin{aligned}
        b & =\frac{\left(-x\pm \sqrt{x^2-4 a f}\right) (f \nu +g x)+2 a f g}{2 f^2}\\ 
        c &=\frac{(x-\mu ) \left(\left(x^2\pm x \sqrt{x^2-4 a f}\right)-2 a f\right)}{2 (f \nu +g x)}\\
        h & =\frac{f^2 (\mu -x)}{f \nu +g x}
    \end{aligned}
    \label{eq:single-flat-gen-sol-1}
\end{equation} 
Then the band structure is 
\begin{equation}
    \begin{aligned}
        \efb &= x\\ 
        E_k &= \mu-x +\frac{\left(a e^{i k}+f e^{-i k}\right) (f \nu +g \mu )}{f \nu +g x } \\
        & -\frac{e^{i k} \nu  (x -\mu ) \left(x \pm \sqrt{x^2-4 a f}\right)}{2 (f \nu +g x )} \; .
    \end{aligned}
    \label{eq:single-flat-gen-band-1}
\end{equation}

Putting corresponding values of $\mu,\nu$ into Eq.~\eqref{eq:single-flat-gen-sol-1} and~\eqref{eq:single-flat-gen-band-1}, we can get solutions for degenerate, non-degenerate and abnormal cases and the corresponding band structures.

\section{Solving partial flatbands}
\label{app:partial-fb} 

We assume that $x_k=x_1 + i x_2, y=y_1 + i y_2$, and $x_1,x_2,y_1,y_2 \in \mathcal{R}$, then complex expanding equation~\eqref{eq:band-eqn-app} yields
\begin{equation}
    \begin{aligned} 
        x_k + y_k &= x_{1} +y_{1} +i\left( x_{2} + y_{2} \right) = \mu \\
        & +\cos(k)(a+d+f+l)\\
        & -i\sin(k)(a+d-f-l)\\ 
        x_k y_k &= x_1 y_1 - x_2 y_2 + i\left(x_{2}y_{1}+x_{1}y_{2}\right) \\
        &= al - bh - cg + df\\
        & +(a\mu - c\nu + f\mu - h\nu)\cos(k)\\
        & +(\det H_l + \det H_r)\cos(2k)\\
        & +i(\left(-a\mu + c\nu + f\mu - h\nu\right)\sin(k)\\
        & +(\det H_l - \det H_r)\sin(2k)) \; .
    \end{aligned}
    \label{eq:re-im-gen-eq}
\end{equation}
Equating real and imaginary parts of Eq.~\eqref{eq:re-im-gen-eq} 
\begin{align}
    \label{eq:complex-exp-band-eq-1}
    x_1 + y_1 &=& \mu + (a + d + f + l)\cos(k)\\
    \label{eq:complex-exp-band-eq-1-1}
    x_2 + y_2 &=& -(a + d - f - l)\sin(k)\\
    x_1 y_1 - x_2 y_2 &=& al - bh - cg + df\notag\\
    \label{eq:complex-exp-band-eq-1-2}
    && +(a\mu-c\nu+f\mu-h\nu)\cos(k)\\ 
    &&+(\det H_{l}+\det H_{r})\cos(2k)  \notag\\
    \label{eq:complex-exp-band-eq-2}
    x_{2}y_{1}+x_{1}y_{2}&=&\left(-a\mu+c\nu+f\mu- h\nu\right)\sin(k)\\ 
    &&  +(\det H_l - \det H_r)\sin(2k)  \notag \; .
\end{align}
Solving equations~(\ref{eq:complex-exp-band-eq-1}-\ref{eq:complex-exp-band-eq-2}) under the condition that some of $x_1,x_2,y_1,y_2$ to be $k$-independent, we get the solution for partially flatbands. 
  
\subsection{Real parts of both bands are flat}
\label{app:both-re-fb}

In this case $x_1,y_1$ are $k$ independent, then Eqs.~(\ref{eq:complex-exp-band-eq-1}-\ref{eq:complex-exp-band-eq-1-1}) gives 
\begin{equation}
    \begin{aligned}
        y_1 &= \mu- x_1 + (a+d+f+l)\cos(k)\\
        y_2 &= -x_2-(a+d-f-l)\sin(k)\; .
    \end{aligned}
    \label{eq:both-re-fb-y1-y2}
\end{equation}
We put above Eq.~\eqref{eq:both-re-fb-y1-y2} into Eqs.~(\ref{eq:complex-exp-band-eq-1-2}-\ref{eq:complex-exp-band-eq-2}), and solve for $x_2$. Then requiring $x_1$ to be $k$-independent, by which making coefficients of $k$-independent terms to be zero, we got the following equations 
\begin{equation}
    \begin{aligned}
        a+d+f+l&= 0 \\ 
        a - c + f - h & =0 \\ 
        a d+f l -b c-g h & =\frac{\left(X + Y \right) \left( X + Z \right) }{2 (\mu -2 x_1)^2}\\
        -a d+b c+f l-g h & =0  \; ,
    \end{aligned}
    \label{eq:both_real_fb}
\end{equation}
where $X=x_1 (-a-d+f+l),\ Y=a \mu -c \nu -f \mu +h \nu,\ Z=c \nu +d \mu -h \nu -l \mu$. 

Solving Eqs.~\eqref{eq:both_real_fb} for degenerate, non-degenerate and abnormal cases separately, we got the $H_r,H_l$ that gives real part of both bands are flat.  

\paragraph{Degenerate case} 

In this case $\mu=\nu=0$, and Eqs.~\eqref{eq:both_real_fb} becomes
\begin{equation}
    \begin{aligned}
        a+d+f+l &=0 \\
        \frac{1}{8} (a+d-f-l)^2+b c+g h &=a d+f l \\
        -a d+b c+f l-g h &=0\\
        \frac{1}{8} (a+d-f-l)^2+a l+d f+x_1^2 &=b h+c g
    \end{aligned}
    \label{eq:both-real-degen-eqs}
\end{equation}
If we consider $x_1$ as a parameter, then the solution for Eq.~\eqref{eq:both-real-degen-eqs} is 
\begin{align*}
    H_0 &= \begin{pmatrix}
        0 & 0\\
        0 & 0
    \end{pmatrix},\\
    H_l &= \begin{pmatrix}
        -f & \frac{\pm 2 \sqrt{A} + B}{(a-d)^2}\\
        -\frac{\pm 2 \sqrt{A} + B}{4 b^2} & -a-d-f
    \end{pmatrix},\\
    H_r &= \begin{pmatrix}
        a & b\\
        -\frac{(a-d)^2}{4 b} & d
    \end{pmatrix}\; .
\end{align*}
where $A=b^2 x_1^2 \left((d-a) (a+d+2 f)+x_1^2\right),\ B=b (a-d) (a+d+2 f)-2 b x_1^2$. 
Then the band structure is 
\begin{equation}
    \begin{aligned}
        x_k = -x_1+i (a+d) \sin (k)\\
        y_k = x_1+i (a+d) \sin (k)
    \end{aligned}
\end{equation}

If we consider $x_1$ as a function of $H_r,H_l$, then the solution for Eqs.~\eqref{eq:both-real-degen-eqs} is
\begin{equation}
    \begin{aligned}
        H_0 = \begin{pmatrix}
            0 & 0\\
            0 & 0
        \end{pmatrix},\\
        H_l = \begin{pmatrix}
            f & -\frac{(a+d+2 f)^2}{4 h} \\
            h & -a-d-f 
        \end{pmatrix},\\
        H_r = \begin{pmatrix}
            a & b \\
            -\frac{(a-d)^2}{4 b} & d \\
        \end{pmatrix},\\
         x_1 = \pm \frac{(a-d) (a+d+2 f)+4 b h}{4 \sqrt{b} \sqrt{h}}\; .
    \end{aligned}
\end{equation}
The band structure is 
\begin{equation}
    \begin{aligned}
        x_k &= -\frac{\sqrt{b h ((a-d) (a+d+2 f)+4 b h)^2}}{4 b h}  \\ & +\frac{4 i b h (a+d) \sin (k)}{4 b h} ,\\
        y_k &= \frac{\sqrt{b h ((a-d) (a+d+2 f)+4 b h)^2}}{4 b h} \\ & + \frac{4 i b h (a+d) \sin (k)}{4 b h}
    \end{aligned}
\end{equation}

\paragraph{Non-degenerate case} 

In this case $\mu=1,\nu=0$, and Eqs.~\eqref{eq:both_real_fb} become
\begin{equation}
    \begin{aligned}
        & a+d+f+l= 0 \\ 
        & a + f =0 \\ 
        & \frac{\left(X+a-f\right) \left(X+d-l\right)}{2 \left(1-2 x_1\right){}^2}+b c+g h=a d+f l\\ 
        & -a d+b c+f l-g h = 0\\
        & \frac{ x_1\left(1- x_1\right)(a-d-f+l)^2}{2 \left(1-2 x_1\right)^2}=-(b h+c g+x_1) \\ 
        & +a l+d f+x_1^2  \; ,
    \end{aligned}
\end{equation} 
where $X=x_1 (-a-d+f+l)$. Then the solution is 
\begin{equation}
    \begin{aligned}
        H_0 &= \begin{pmatrix}
            0 & 0\\
            0 & 1
        \end{pmatrix},\\
        H_r &= \begin{pmatrix}
            -f & -\frac{\left(x_1-1\right) x_1 (f-l)^2}{c\left(1-2 x_1\right){}^2}\\
            c & -l
        \end{pmatrix},\\
        H_l &= \begin{pmatrix}
            f & \frac{\left(x_1-1\right) x_1 (C+D)}{2 c \left(1-2 x_1\right){}^2}\\
            \frac{c (C-D)}{2 (f-l)^2} & l
        \end{pmatrix},
    \end{aligned}
\end{equation}
where $C=\sqrt{4(f-l)^2+(1-2x_1)^2} \left(1-2 x_1\right),\ D=2 (f-l)^2 + (1 - 2x_1)^2$.

Then the band structure is 
\begin{equation} 
    \begin{aligned}
        x_k & = x_1-\frac{2 i \sin (k) \left(x_1 (f+l)-f\right)}{2 x_1-1}\\
        y_k & = -\frac{2 i \sin (k) \left(x_1 (f+l)-l\right)}{2 x_1-1}-x_1+1 \; .
    \end{aligned}
\end{equation}
Obviously the real parts $x_1,1-x_1$ are $k$ independent, i.e flat. 

\paragraph{Abnormal case}

In this case $\mu=0,\nu=1$, and Eqs.~\eqref{eq:both_real_fb} becomes     
\begin{equation}
    \begin{aligned}
        & a+d+f+l =0 \\
        & -c - h = 0 \\
        & \frac{\left(X+c-h\right) \left(X-c+h\right)}{8 x_1^2}  = a d + f l - b c - g h\\
        & -a d+b c+f l-g h  = 0 \\
        & b h+c g+\frac{(c-h)^2}{8 x_1^2} =\frac{1}{8} (a+d-f-l)^2 \\ & +a l+d f+x_1^2  \; ,
    \end{aligned} 
\end{equation} 
where $X=x_1 (-a-d+f+l)$. Then the solution is 
\begin{equation}
    \begin{aligned}
        H_{0}&=\begin{pmatrix}
            0 & 1\\
            0 & 0
        \end{pmatrix},\\
        H_{l}&=\begin{pmatrix}
            f & \frac{h^2-x_1^2 \left(d+f-x_1\right){}^2}{4 h x_1^2}\\
            h & x_1-d
        \end{pmatrix},\\
        H_{r}&=\begin{pmatrix}
            -f-x_1 & \frac{x_1^2 \left(d+f+x_1\right){}^2-h^2}{4 h x_1^2}\\
            -h & d
        \end{pmatrix}\; ,
    \end{aligned}
\end{equation} 
Then the band structure is 
\begin{equation}
    \begin{aligned}
        x_k & = -x_1+i \sin (k) \left(d-f+\frac{h}{x_1}-x_1\right)\\
        y_k & = x_1-i \sin (k) \frac{ \left(x_1 \left(-d+f+x_1\right)+h\right)}{x_1} \; .
    \end{aligned}
\end{equation} 

\subsection{Real part of one band is flat} 

In this case either $x_1$ or $y_1$ is $k$ independent in Eqs.~(\ref{eq:complex-exp-band-eq-1}-\ref{eq:complex-exp-band-eq-2}). If we assume $x_1$ is $k$ independent and solve Eqs.~(\ref{eq:complex-exp-band-eq-1}-\ref{eq:complex-exp-band-eq-2}) for $x_1$, then according to Eq.~\eqref{eq:complex-exp-band-eq-1} we have 
\begin{equation}
    \begin{aligned}
        x_1 &= \mu \\
        y_1 &= (a+d+f+l)Cos (k) 
    \end{aligned}
\end{equation}
(In similarly way we can assume $y_1$ is $k$ independent, then $y_1=\mu, x_1=(a+d+f+l)Cos (k)$.) Then Eqs.~(\ref{eq:complex-exp-band-eq-1}-\ref{eq:complex-exp-band-eq-2}) becomes 
\begin{equation}
    \begin{aligned}
        x_2+y_2 &= \sin (k) (a+d-f-l) \\ 
        x_2 y_2 &= -a l+b h+\cos (k) (\nu  (c+h)+d \mu +l \mu ) \\ 
        & +c g-d f-(\det H_l + \det H_r) \cos (2 k) \\
        \mu  y_2 &= \sin (k) (a \mu -c \nu -f \mu +h \nu ) \\ 
        & -x_2 \cos (k) (a+d+f+l) \\ 
        & +(\det H_r - \det H_l) \sin (2 k)
    \end{aligned}
    \label{eq:band-eq-single-re-flat}
\end{equation}
For convenience we make the following replacement of variables 
\begin{equation}
    \begin{aligned}
        \det H_r &= a d-b c\\
        \det H_l &= f l-g h\\
        V &= a+d+f+l\\
        X &= a+d-f-l\\
        Y &= a \mu -c \nu +f \mu -h \nu\\
        Z &= a \mu -c \nu -f \mu +h \nu \\
        W &= a l-b h-c g+d f
    \end{aligned}
    \label{eq:rep-var}
\end{equation}
Then Eqs.~\eqref{eq:band-eq-single-re-flat} becomes 
    \begin{align}
        x_1+y_1 &=V \cos (k)+\mu \label{eq:single-re-fb-1}\\ 
        x_2+y_2 &=X \sin (k) \label{eq:single-re-fb-2}\\ 
        x_1 y_1-x_2 y_2 &= (\det H_l+\det H_r) \cos (2 k) \label{eq:single-re-fb-3 }\\ 
        & + Y \cos (k) +W \notag\\
        x_2 y_1+x_1 y_2 &=-(\det H_l-\det H_r)\sin (2k) \label{eq:single-re-fb-4} \\ 
        & + Z \sin (k) \notag
    \end{align}
We can solve Eq.~\eqref{eq:single-re-fb-4} three variables: $x_2,y_2$ and third variable is one of the $X,Y,Z,U,V,W$. The we require the third variable to be $k$ independent by zeroing coefficients  of all $k$ dependent terms. This gives a set of equation, and solving it gives the solution for real part of one band is flat case. Following are our results.

\paragraph{Degenerate cases:} In this case $\mu=0,\nu=0$, and the solution is
\begin{equation}
    \begin{aligned}
        H_r & = \left(
        \begin{array}{cc}
            a & b \\
            \frac{(a+f) (b (d+f)+(d-a) g)}{(b+g)^2} & d \\
        \end{array}
        \right), \\
        H_l &= \left(
            \begin{array}{cc}
            f & g \\
            \frac{(a+f) (g (a+l)+b (l-f))}{(b+g)^2} & l \\
            \end{array}
        \right) \; .
    \end{aligned}
    \label{eq:single-re-fb-degen-sol}
\end{equation} 
The band structure is
\begin{equation}
    \begin{aligned}
        E_1 &= -\frac{2 i \sin (k) (b f-a g)}{b+g}\\ 
        E_2 &= \frac{\left(e^{ i k} (b (a+d+f)+d g)\right)}{b+g} \\ 
        & + \frac{\left(e^{-i k}(a g+b l+f g+g l)\right)}{b+g}
    \end{aligned}
\end{equation} 

\paragraph{Abnormal case:} In this case $\mu=0,\nu=1$, and the solution is 
\begin{equation}
    H_r = \left(
    \begin{array}{cc}
        a & b \\
        0 & d \\
    \end{array}
    \right), \quad H_l = \left(
    \begin{array}{cc}
        -a & g \\
        0 & l \\
    \end{array}
    \right)
    \label{eq:single-re-fb-abn-sol}
\end{equation} 
The band structure is 
\begin{equation}
    \begin{aligned}
        E_1 &= 2 i a \sin (k)\\ 
        E_2 &= d e^{i k}+e^{-i k} l+1
    \end{aligned}
\end{equation}
Note that when $a=-f$, the solution~\eqref{eq:single-re-fb-degen-sol} for degenerate case reduces to abnormal case~\eqref{eq:single-re-fb-abn-sol}. 

\paragraph{Non-degenerate case:} In this case $\mu=1,\nu=0$, and the solution is
\begin{equation}
    H_r = \left(
    \begin{array}{cc}
        a & 0 \\
        c & d \\
    \end{array}
    \right), \quad H_l = \left(
    \begin{array}{cc}
        f & 0 \\
        h & -d \\
    \end{array}
    \right)
\end{equation} 
The band structure is
\begin{equation}
    \begin{aligned}
        E_1 &= 1+2 i d \sin (k)\\ 
        E_2 &= e^{-i k} \left(f+a e^{2 i k}\right)
    \end{aligned}
\end{equation} 

\subsection{Imaginary part of both bands are flat}
\label{app:both-im-fb}

In this case $x_2,y_2$ are $k$ independent in Eqs.~(\ref{eq:complex-exp-band-eq-1}-\ref{eq:complex-exp-band-eq-2}). Using the same procedure as in Sec.~\ref{app:both-re-fb}, solving Eqs.~(\ref{eq:complex-exp-band-eq-1}-\ref{eq:complex-exp-band-eq-2}) for $y_2$, and requiring $y_1$ to be $k$ independent, we find 
\begin{equation}
    \begin{aligned}
        a+d & = f+l\\
        b c+f l &= a d+g h\\
        16 x_2^2 (b c-a d) &= -(a \mu +\nu  (h-c)-f \mu )^2  \\ 
        & - x_2^2(a+d+f+l)^2\\
        \mu  x_2 (a+d+f+l) &= 2 x_2 (a \mu -\nu  (c+h)+f \mu )\\
        8 x_2^4 + 2 \mu^2 x_2^2 &= 8 x_2^2 (a l-b h-c g+d f) \\ 
        & - x_2^2 (a+d+f+l)^2 \\ 
        & +(a \mu +\nu  (h-c)-f \mu )^2
    \end{aligned}
    \label{eq:both_im_fb_cond}
\end{equation} 

\paragraph{Degenerate case} 

In this case $\mu=\nu=0$, and Eq.~\eqref{eq:both_im_fb_cond} becomes 
\begin{equation}
    \begin{aligned}
        a+d &=f+l\\
        b c+f l &=a d+g h\\
        16 x_2^2 (b c-a d) &= -x_2^2(a+d+f+l)^2\\
        x_2^2(a+d+f+l)^2 &= 8 x_2^2(a l-b h-c g+d f)\\ & -8 x_2^4
    \end{aligned}
\end{equation}
The solution is 
\begin{equation}
    \begin{aligned}
        H_{0}&=\begin{pmatrix}
            0 & 0\\
            0 & 0
        \end{pmatrix},\\
        H_{l}&=\begin{pmatrix}
            f & \frac{b \left(F+2 x_2^2\right)}{(a-d)^2}\\
            \frac{F-2 x_2^2}{4 b} & a+d-f
        \end{pmatrix},\\
        H_{r}&=\begin{pmatrix}
            a & b\\
            -\frac{(a-d)^2}{4 b} & d
        \end{pmatrix}\; ,
    \end{aligned}
\end{equation}
where $F=-2 \sqrt{x_2^2 (d-a) (a+d-2 f)+x_2^4}+(d-a) (a+d-2 f)$. Then the band structure is
\begin{equation}
    \begin{aligned}
        x_k &= \frac{ 2 b (a-d)^2 (a+d)\cos k }{2 b (a-d)^2}\\
        &- \frac{ 2 \sqrt{-b^2 x_2^2 (a-d)^4}}{2 b (a-d)^2}\\
        y_k &= \frac{2 b (a+d) (a-d)^2 \cos k}{2 b (a-d)^2} \\
        &+ \frac{2 \sqrt{-b^2 x_2^2 (a-d)^4}}{2 b (a-d)^2}
    \end{aligned}
\end{equation} 

\paragraph{Non-degenerate case:} In this case $\mu=1,\nu=0$, and Eq.~\eqref{eq:both_im_fb_cond} becomes 
\begin{equation}
    \begin{aligned}
        & a+d=f+l\\
        & b c+f l=a d+g h\\
        & x_2^2 \left(16 (b c-a d)+(a+d+f+l)^2\right)+(a-f)^2=0\\
        & x_2 (a-d+f-l)=0\\
        & x_2^2 \left(8 (a l-b h-c g+d f)-(a+d+f+l)^2-2\right)\\ &+(a-f)^2=8 x_2^4
    \end{aligned}
\end{equation}
The solution is 
\begin{equation}
    \begin{aligned}
        H_{0}&=\begin{pmatrix}
            0 & 0\\
            0 & 1
        \end{pmatrix},\\
        H_{l}&=\begin{pmatrix}
                f & \frac{-G}{(a-f)^2}\\
                \frac{\left(4 x_2^2+1\right) (a-f)^4}{16 x_2^2 G} & a
        \end{pmatrix},\\
        H_{r}&=\begin{pmatrix}
            a & b\\
            -\frac{\left(4 x_2^2+1\right) (a-f)^2}{16 b x_2^2} & f
        \end{pmatrix}\; .
    \end{aligned}
\end{equation} 
where $G=2 \sqrt{b^2 x_2^2 \left(x_2^2-(a-f)^2\right)}+b (a-f)^2-2 b x_2^2$. Then the band structure is 
\begin{equation}
    \begin{aligned}
        x_k & =-\frac{K}{4 b G x_2^2 (a-f)^2}+(a+f) \cos (k)+\frac{1}{2}\\
        y_k &= \frac{K}{4 b G x_2^2 (a-f)^2}+(a+f) \cos (k)+\frac{1}{2}
    \end{aligned}
\end{equation}
where 
\begin{equation*}
    \begin{aligned}
        K &= \Bigg[b G x_2^2 (a-f)^4 \Bigg(4 x_2^2 \Big(b^2 (a-f)^4-4 i b G (a-f) \sin (k) \\ 
        & +b G \left(1-2 (a-f)^2\right)+G^2\Big)+b^2 (a-f)^4 \\ 
        &-2 b G (a-f)^2 \cos (2 k)+G^2\Bigg)\Bigg]^{\sfrac{1}{2}}
    \end{aligned}
\end{equation*} 

\paragraph{Abnormal case}

In this case $\mu=0,\nu=1$, and Eq.~\eqref{eq:both_im_fb_cond} becomes 
\begin{equation}
    \begin{aligned}
        a+d &=f+l\\
         b c+f l &=a d+g h\\
         (c-h)^2 &=-16 x_2^2  (b c-a d)\\ 
         &-x_2^2 (a+d+f+l)^2\\
         x_2 (c+h) &=0\\ 
         (c-h)^2 - 8 x_2^4 &= x_2^2 (a+d+f+l)^2 \\ 
         & - 8 x_2^2 (a l-b h-c g+d f)
    \end{aligned}
\end{equation}
The solution is 
\begin{equation}
    \begin{aligned}
        H_{0}&=\begin{pmatrix}
            0 & 1\\
            0 & 0
        \end{pmatrix},\\
        H_{l}&=\begin{pmatrix}
            d-i x_2 & \frac{c}{4 x_2^2}-\frac{x_2^2}{c}\\
            -c & d+i x_2
        \end{pmatrix},\\
        H_{r}&=\begin{pmatrix}
            d & -\frac{c}{4 x_2^2}\\
            c & d
        \end{pmatrix}\; .
    \end{aligned}
\end{equation}
The band structure is
\begin{equation}
    \begin{aligned}
       x_k &= 2 d \cos k -\frac{ \sqrt{c^2 x_2^2 \left(c \sin (k)+i x_2^2\right){}^2}}{c x_2^2} \\
       y_k &= 2 d \cos k + \frac{ \sqrt{c^2 x_2^2 \left(c \sin (k)+i x_2^2\right){}^2}}{c x_2^2} 
    \end{aligned}
\end{equation} 

\subsection{Imaginary part of one band is flat}
\label{app:single-im-fb}

Using the same method with real part of one band is flat case, we can also solve the case of imaginary part of one band is flat case. In this case we require $x_2$ ($y_2$) to be $k$ independent in Eqs.~(\ref{eq:complex-exp-band-eq-1}-\ref{eq:complex-exp-band-eq-2}). The only possibility is $x_2=0, y_2=(a+d-f-l)$, then following same steps with real part of one band case we got the following results. 

\paragraph{Degenerate case:}

\begin{equation}
    \begin{aligned}
        H_r &= \begin{pmatrix}
            a & b\\
            \frac{(a-f) (g (a-d)+b (d-f))}{(b-g)^2} & d
        \end{pmatrix}, \\ 
        H_l &= \begin{pmatrix}
            f & g\\
            \frac{(a-f) (g (a-l)+b (l-f))}{(b-g)^2} & l
        \end{pmatrix}
    \end{aligned}
\end{equation} 
The band structure is 
\begin{equation}
    \begin{aligned}
        E_1 &= \frac{2 \cos (k) (b f-a g)}{b-g} \\ 
        E_2 &= \frac{ e^{ i k} (b (a+d-f)-d g)}{b-g}\\
        &+\frac{ e^{-i k} (a g+b l-f g-g l)}{b-g}
    \end{aligned}
\end{equation} 

\paragraph{Abnormal case:}
\begin{equation}
    H_r = \left(
    \begin{array}{cc}
        a & b \\
        0 & d \\
    \end{array}
    \right), \quad 
    H_l = \left(
    \begin{array}{cc}
        a & g \\
        0 & l \\
    \end{array}
    \right)
\end{equation} 
The band structure is 
\begin{equation}
    \begin{aligned}
        E_1 &= 2 a \cos (k) \\ 
        E_2 &= e^{-i k} \left(l+d e^{2 i k}\right)
    \end{aligned}
\end{equation} 

\paragraph{Non-degenerate case}
In this case the solution is the same with abnormal case, and the band structure is 
\begin{equation}
    \begin{aligned}
        E_1 &= 2 a \cos (k) \\
        E_2 &= d e^{i k}+e^{-i k} l+1
    \end{aligned}
\end{equation} 

\subsection{Modulus of a band is flat}
\label{app:mod-fb}

We suppose $x=r e^{i\theta_{k}}$ in Eq.~\eqref{eq:bands-eqn-allf-xy}, then 
\begin{equation}
    \begin{aligned}
        y & = e^{ik}(a+d)+e^{-ik}(f+l)+\mu-re^{i\theta_{k}} \\
        y & = \frac{1}{r}e^{-i\theta_{k}}\Big(e^{ik}(a\mu-c\nu)+al-bh-cg+df \\
        & +\det H_{l}e^{-2ik}+\det H_{r}e^{2ik}+e^{-ik}(f\mu-h\nu)\Big)
    \end{aligned}
\end{equation} 
Since there only integer powers of $e^{i k}$ present in the above, $\theta_k = m k ,\ m \in \mathcal{Z}$ and 
\begin{equation}
    \begin{aligned}
        & e^{i\left(m-1\right)k}(f+l) +e^{i\left(m+1\right)k}(a+d) \\
        & +\mu e^{imk} -re^{2imk} \\ 
        &=\frac{1}{r} \Big(e^{-ik}(f\mu-h\nu) \\ & +al-bh-cg+df+e^{ik}(a\mu-c\nu) \\ 
        & + \det H_{l}e^{-2ik} +\det H_{r}e^{2ik} \Big)
    \end{aligned}
   \label{eq:generic-m}
\end{equation} 
By equating same powers of $e^{i k}$ in Eq.~\eqref{eq:generic-m}, we can show that, when $m>1$ or $m<-1$ only possible solution for Eq.~\eqref{eq:generic-m} is $r=0$. Therefore, only when $m=0, \pm 1$ we have non-trivial solution. Note that $m=0$ case corresponds to one band is completely flat, which is solved in Appendix~\ref{app:one-band-fb}.

\subsubsection{m=1 case} 

In this case equating the coefficients of the same powers of $e^{ik}$ on two sides of Eq.~\eqref{eq:generic-m} we get 
\begin{equation}
    \begin{aligned}
        f+l & =\frac{1}{r}\left(al-bh-cg+df\right)\\
        \mu & =\frac{1}{r}\left(a\mu-c\nu\right)\\
        a+d-r & =\frac{1}{r}\det H_{r}\\
        \det H_{l} & =0\\
        f\mu-h\nu & =0
    \end{aligned}
    \label{eq:mod-fb-m1-eqs}
\end{equation} 

\paragraph{Degenerate case:} In this case $\mu=0, \nu=0$, and Eq.~\eqref{eq:mod-fb-m1-eqs} gives following solution 
\begin{equation}
    \begin{aligned}
        H_{r} & =\left(
        \begin{array}{cc}
            a & b\\
            -\frac{(a-r)(r-d)}{b} & d
        \end{array}\right)\\
        H_{l} & =\left(
        \begin{array}{cc}
            f & \frac{bf}{a-r}\\
            \frac{l(a-r)}{b} & l
        \end{array}\right) \; ,
    \end{aligned}  
\end{equation}
Then the band structure is 
\begin{equation}
    \begin{aligned}
        E_{1} & =e^{ik}r,\\
        E_{2} & =e^{-ik}\left(ae^{2ik}+de^{2ik}+f-e^{2ik}r+l\right)
    \end{aligned}
\end{equation} 

\paragraph{Non-degenerate case:} In this case $\nu=0,\mu=1$, and Eq.~\eqref{eq:mod-fb-m1-eqs} gives following solution 
\begin{equation}
    \begin{aligned}
        H_{r} & =\left(
        \begin{array}{cc}
            r & 0\\
            c & d
        \end{array}\right)\\
        H_{l} & =\left(
        \begin{array}{cc}
            0 & 0\\
            h & l
        \end{array}\right) \; ,
    \end{aligned}  
\end{equation}
band structure is 
\begin{equation}
    \begin{aligned}
        E_{1} & =e^{ik}r,\\
        E_{2} & =de^{ik}+e^{-ik}l+1
    \end{aligned}  
\end{equation} 

\paragraph{Abnormal case:} In this case $\mu=0,\nu=1$, and Eq.~\eqref{eq:mod-fb-m1-eqs} gives following solution 
\begin{equation}
    \begin{aligned}
        H_{r} & =\left(
        \begin{array}{cc}
            r & b\\
            0 & d
        \end{array}\right)\\
        H_{l} & =\left(
        \begin{array}{cc}
            0 & g\\
            0 & l
        \end{array}\right)
    \end{aligned}
\end{equation} 
The band structure is 
\begin{equation}
    \begin{aligned}
        E_{1} & =e^{ik}r,\\
        E_{2} & =de^{ik}+e^{-ik}l+1
    \end{aligned}  
\end{equation} 

\subsubsection{ m=-1 case} 

Similar to $m=1$ case, by equating the same powers of $e^{i k}$ in Eq.~\eqref{eq:generic-m}, we have 
\begin{equation}
    \begin{aligned}
        f+l-r & =\frac{1}{r}\det H_{l}\\
        \frac{1}{r}\left(f\mu-h\nu\right) & =\mu\\
        a+d & =\frac{1}{r}\left(al-bh-cg+df\right)\\
        \det H_{r} & =0\\
        a\mu-c\nu & =0
    \end{aligned} 
\end{equation} 

\paragraph{Degenerate case:} In this case $\mu=0, \nu=0$, and Eq.~\eqref{eq:generic-m} gives the following solution 
\begin{equation}
    \begin{aligned}
        H_{r} & =\left(\begin{array}{cc}
        a & b\\
        \frac{ad}{b} & d
        \end{array}\right)\\
        H_{l} & =\left(\begin{array}{cc}
        f & \frac{b(f-r)}{a}\\
        \frac{a(l-r)}{b} & l
        \end{array}\right)
    \end{aligned}
\end{equation}
 Then the band structure is 
\begin{equation}
    \begin{aligned}
        E_{1} & =e^{-ik}r,\\
        E_{2} & =e^{-ik}\left(ae^{2ik}+de^{2ik}+f+l-r\right)
    \end{aligned}  
\end{equation}

\paragraph{Non-degenerate case:} In this case $\nu=0,\mu=1$, and Eq.~\eqref{eq:generic-m} gives the following solution
\begin{equation}
    \begin{aligned}
        H_{r} & =\left(
        \begin{array}{cc}
            0 & 0\\
            c & d
        \end{array}\right)\\
        H_{l} & =\left(
        \begin{array}{cc}
            r & 0\\
            h & l
        \end{array}\right)
    \end{aligned}
\end{equation}
 Then the band structure is 
\begin{equation}
    \begin{aligned}
        E_{1} & =e^{-ik}r,\\
        E_{2} & =de^{ik}+e^{-ik}l+1
    \end{aligned}
\end{equation} 

\paragraph{Abnormal case:} In this case $\nu=1,\mu=0$, and Eq.~\eqref{eq:generic-m} gives the following solutions 
\begin{equation}
    \begin{aligned}
        H_{r} & =\left(
        \begin{array}{cc}
            0 & b\\
            0 & d
        \end{array}\right)\\
        H_{l} & =\left(
        \begin{array}{cc}
            r & g\\
            0 & l
        \end{array}\right)
    \end{aligned}
\end{equation}
Then the band structure is 
\begin{equation}
    \begin{aligned}
        E_{1} & =e^{-ik}r,\\
        E_{2} & =e^{-ik}\left(l+de^{2ik}\right)
    \end{aligned}
\end{equation}

\bibliography{general,flatband}

\end{document}